\documentclass[usenatbib,useAMSm a4paper, singlespacing]{mn2e}

 \usepackage{caption}
\usepackage{subcaption}
 
\usepackage{graphicx}	
\usepackage{amsmath}	
\usepackage{amssymb}	
\usepackage{multirow,multicol}        
\usepackage{bm}		
\usepackage{pdflscape}
\usepackage[T1]{fontenc}
\usepackage{txfonts}

\usepackage{longtable,lscape, booktabs,etoolbox}
\usepackage{float}

\usepackage{rotating}
\usepackage{epsfig}
\usepackage{footnote}
\usepackage[referable]{threeparttablex}
\usepackage{hyperref} 
\usepackage{bookmark}

\def\mearth{\ifmmode {\rm M_{\oplus}}\else $\rm M_{\oplus}$\fi}
\def\Mearth{\ifmmode {\rm M_{\oplus}}\else $\rm M_{\oplus}$\fi}
\def\Rearth{\ifmmode {\rm R_{\oplus}}\else $\rm R_{\oplus}$\fi}
\def\Ms{\ifmmode {M_s}\else $M_s$\fi}
\def\Mp{\ifmmode {M_p}\else $M_p$\fi}
\def\Rp{\ifmmode {R_p}\else $R_p$\fi}
\def\rearth{\ifmmode {\rm R_{\oplus}}\else $\rm R_{\oplus}$\fi}

\def\aap{A\&A}

\def\apjl{ApJL}
\def\apjs{ApJS}

\newcommand{\Msun}{M_{\odot}}

\newcommand{\Rsun}{R_\odot}

\newcommand{\Lsun}{L_{\odot}}

\title[The Very Wide Outflow Driven by HBC 494 ]{The ALMA Early Science View of FUor/EXor objects. II. The Very Wide Outflow Driven by HBC 494 \thanks{Based on ALMA observations, program number 2013.1.00710.S}}

\author[D. Ru\'iz-Rodr\'iguez et al.]
{ \Large D. Ru\'iz-Rodr\'iguez, $^1$\thanks{E-mail:dary.ruiz@anu.edu.au}
L. A. Cieza,$^{2,3}$
J. P. Williams,$^4$
J. J. Tobin,$^5$ 
A. Hales,$^6$ 
Z. Zhu,$^{7,8}$
K. Mu\v{z}i\'c,$^3$
\newauthor \Large D. Principe,$^{2,3}$
H. Canovas,$^{9}$ 
A. Zurlo,$^{2,3,10}$
S. Casassus,$^{2,10}$ 
S. Perez,$^{2,10}$ and J. L. Prieto $^3$
\\
$^{1}$Research School of Astronomy and Astrophysics, Australian National University, Canberra, ACT 2611, Australia\\
$^{2}$Millenium Nucleus ``Protoplanetary discs in ALMA Early Science", Chile \\
$^{3}$N\'ucleo de Astronom\'ia,  Facultad de Ingenier{\'i}a, Universidad Diego Portales,  Av. Ejercito 441, Santiago, Chile\\
$^{4}$Institute for Astronomy, University of Hawaii at Manoa, Honolulu, HI, 96822, USA\\
$^{5}$Homer L. Dodge Department of Physics and Astronomy, University of Oklahoma, 440 W. Brooks Street, Norman, OK 73019, USA\\
$^{6}$Atacama Large Millimeter/Submillimeter Array, Joint ALMA Observatory, Alonso de C\'ordova 3107, Vitacura 763-0355, Santiago, Chile\\
$^{7}$Department of Astrophysical Sciences, Princeton University, Princeton, NJ 08544, USA\\
$^{8}$Department of Physics and Astronomy, University of Nevada, Las Vegas, 4505 South Maryland Parkway, Las Vegas, NV 89154, USA\\
$^{9}$Departamento de F\'isica Te\'orica, Universidad Aut\'onoma de Madrid, Cantoblanco 28049 Madrid, Spain\\
$^{10}$Universidad de Chile, Camino el Observatorio 1515, Santiago, Chile
}

\date{}
\pubyear{2016}

\begin{document}

\label{firstpage}
\pagerange{\pageref{firstpage}--\pageref{lastpage}}
\maketitle

\begin{abstract}
We present Atacama Large Millimeter/sub-millimeter Array (ALMA) Cycle-2 observations of the HBC 494 molecular outflow and envelope. HBC 494 is an FU Ori-like object embedded in the Orion A cloud  and is associated with the reflection nebulae Re50 and Re50N. We use $^{12}$CO,  $^{13}$CO  and  C$^{18}$O spectral line data to independently describe the outflow and envelope structures associated with HBC 494.  The moment-1 map of the  $^{12}$CO emission shows the \textit{widest} outflow cavities in a Class I object known to date (opening angle $\sim$ 150$^{^{\circ}}$). The morphology of the wide outflow is likely to be due to the  interaction between winds originating in the inner disc and the surrounding envelope. The low-velocity blue- and red-shifted $^{13}$CO and C$^{18}$O emission trace the rotation and infall motion of the circumstellar envelope. Using molecular line data and adopting standard methods for correcting optical depth effects, we estimate its kinematic properties, including an outflow mass on the order of 10$^{-1}$ M$_{\odot}$. Considering the large estimated outflow mass for HBC 494, our results support recent theoretical work suggesting that wind-driven processes might dominate the evolution of protoplanetary discs via energetic outflows.

\end{abstract}


\section{Introduction}\label{sec: intro}

FU Orionis objects (FUors) belong to an embedded pre-main-sequence phase of young stellar evolution, usually associated with reflection nebulae \citep{Herbig1966, Herbig1977}. Observational features of these objects include similarities to the F-G supergiant optical spectra and overtone CO absorption, in addition to water vapour bands in the near-infrared wavelengths characteristic of K-M supergiants \citep{Mould1978}. At far-infrared/submillimeter wavelengths, the Spectral Energy Distributions  (SED) of  FUors are largely dominated by the envelope emission. Where such envelopes are massive enough to replenish the circumstellar disc in the stellar formation process \citep{Sandell2001}. Usually, these types of young stellar objects (YSO) are explained with a disc in Keplerian rotation that produces the observed double-peaked line profiles as seen in FUors  \citep{Hartmann1985}; however, those objects that do not present line broadening consistent with pure Keplerian rotation might require an additional contribution from another component such as a high-velocity inner disc wind \citep{Eisner2011}.

The main observational feature of FUors is their eruptive variability in optical light that can reach 5 mag or even more. This variability is due to outbursts that rise in short periods of time of around $\sim$ 1$-$10 yrs and decay timescales that take place from decades to centuries \citep{Herbig1966, Herbig1977}. It is believed that material falling from a massive circumstellar disc to the central protostar at high disc accretion rates ($\sim$ 10$^{-4}$ M$_{\odot}$yr$^{-1}$) is responsible for these events \citep{Hartmann1996}, although their outburst frequency is unknown. During the outburst phase, large amounts of disc material ($\sim$ 0.01 M$_{\odot}$) are accreted onto the parent star, thus increasing the luminosity during these short events. If the episodic accretion scenario \citep{Audard2014} is correct, most systems undergo multiple FU Ori events during their evolution and the study of these outbursts represents a key element of the star and planet formation process. 

Additionally, the outbursts might be connected to the evolution and extension of the observed outflows in FUors. It has been suggested that the formation, evolution and widening of molecular outflows are the result of the wide-angle wind that arises from the interaction of the highly accreting disc inner edges with a strongly magnetised central star \citep{Snell1980, Shu2000}; or highly collimated jets that propagate into the surrounding envelope material \citep{Raga1993, Ostriker2001}. However, the triggering mechanism for an FUor outburst has yet to be established. The proposed mechanisms for the FU Ori outburst include: 1) Tidal interaction of a massive disc and an eccentric binary system or a giant planet \citep{Bonnell1992, Lodato2004}, 2)  Magnetorotational instability (MRI) activated by gravitational instabilities (GI) \citep{Armitage2001, Zhu2009, Martin2011} and 3) Disc Fragmentation developing spiral structures, more specifically, clump accretion events \citep{Vorobyov2005}.
Testing the proposed theoretical scenarios requires measuring disc masses of FUor objects and spatially resolving structures such as asymmetries in the disc and close binaries. For instance, clumps would indicate large-scale disc fragmentation, while knowledge of disc mass can constrain if GI operates in these discs.
Interferometric observations in the millimetre and sub-millimetre of FU Orionis objects can spatially and spectrally resolve the envelope, bipolar outflows, and disc emissions and thus, provide a more accurate description of these embedded systems than possible with single-dish observations. 
%

\begin{figure}
\centering
\includegraphics[width=0.4\textwidth]{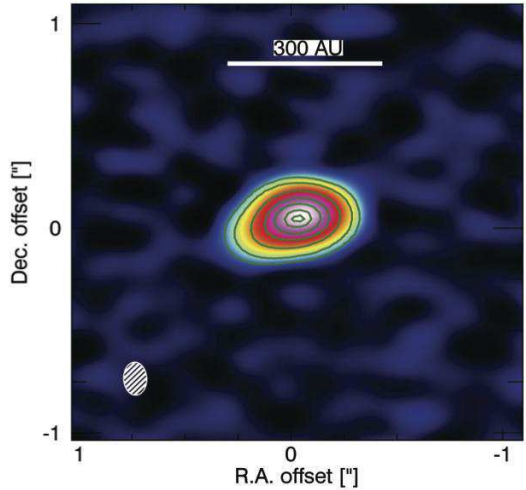}
\caption{Continuum emission of HBC 494 together with contour levels with steps of 10, 30, 80, 150 and 250 $\times$ rms (0.25 mJy beam$^{-1}$). The ${0.35}''\times{0.27}''$with P.A. $=$ -90$^{^{\circ}}$ synthesised beam is shown on the lower left corner of the image. }
\label{Fig:continuum}
\end{figure}

HBC 494 is an FUor object, Class I protostar,  located in the Orion molecular cloud, which has an estimated distance of 414 $\pm$7 pc \citep{Menten2007}. Initially, this luminous object was identified from its associated reflection nebulae, Re50 and Reipurth 50 N (Re50N). \citet{Reipurth1985} detected a very bright nebulous object by performing an optical survey in the Orion molecular cloud. A more detailed study by \citet{Reipurth1986} reported the sudden appearance of an intense and variable conical nebula: Reipurth 50 N (Re50N). This brightening episode is believed to be a consequence of an outburst event in HBC 494. This infrared source has a luminosity of $\sim$ 250 L$_{\odot}$ and is located about 1.5 arc min north of Re50 \citep{ Reipurth1986}. Supporting this hypothesis, \citet{Chiang2015} reported a new brightening event in Re50N at some point between 2006 and 2014, while Re50 has faded considerably. From recent ALMA observations, \citet{Cieza2016} reported an asymmetry in the 230 GHz continuum at the South-West of HBC 494 and they speculated  that this feature might be a result of a  binary object undergoing a formation process and triggering an outburst in HBC 494. Binary objects are becoming strong candidates to trigger these outburst events as they are being resolved more efficiently in the ALMA era, such as the case of the FU Ori system \citep{Hales2015}. Only for guidance to the reader, an illustration of the 230 GHz continuum of HBC 494 is presented in Figure \ref{Fig:continuum} to highlight the non-symmetric emission. From the major and minor axes of the continuum emission, \citet{Cieza2016} found a very high inclination ($i$) of $\sim$ 70.2$^{^{\circ}}$$\pm$ 2.5 (i.e. close edge-on). However, at this resolution, the continuum observation reveals an asymmetry towards the south-east side of the disk, leading to an uncertain estimation of this parameter, a more detailed description and analysis can be found in \citet{Cieza2016}. With a total flux density of 113 $\pm$ 2.5 mJy and adopting a distance to the Orion nebula of 414 $\pm$ 7 pc, they estimated a dust disc mass of 2.0  M$_{J}$ at 20 K,  and assuming a gas-to-dust mass ratio of 100, a total disc mass of  0.2 M$_{\odot}$.

Here, we present ALMA band-6 (230 GHz/1.3 mm) continuum and  $^{12}$CO, $^{13}$CO and  C$^{18}$O  J=2-1 line observations of HBC 494. We use the more optically thin tracers  $^{13}$CO  and C$^{18}$O  to study the envelope material and  the optically thick $^{12}$CO emission to investigate the strong bipolar outflow. The ALMA observations and the data reduction process are described in section \ref{Sec:Observations}. The results are presented in Section \ref{Sec:Results},  and their implications discussed in section \ref{Sec:Discussion}. The summary and conclusion are presented in Section  \ref{Sec:Summary}.

\section[]{Observations}
\label{Sec:Observations}

\begin{figure*}
\centering
\includegraphics[width=0.6\textwidth]{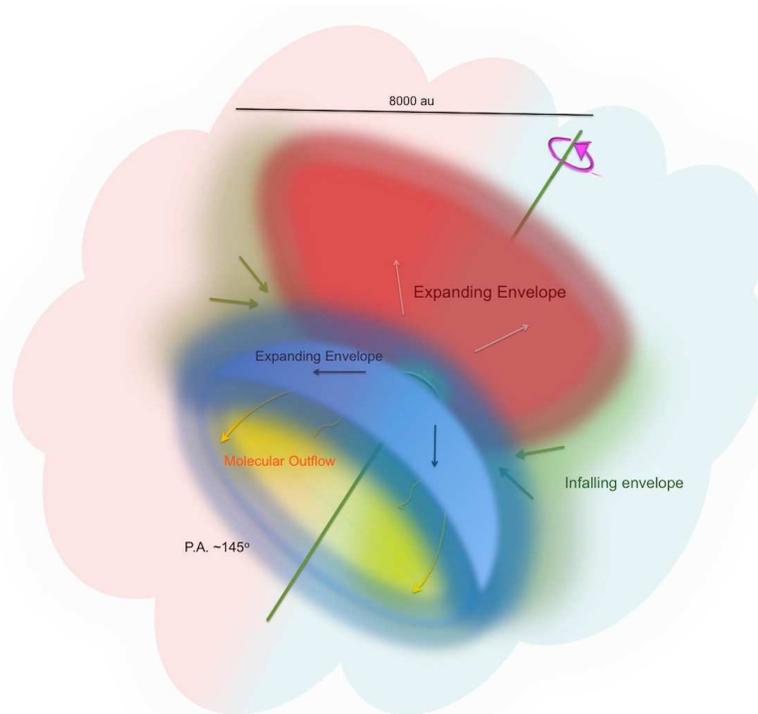}
\caption{Cartoon showing the different dynamical and flux components traced by $^{12}$CO, $^{13}$CO and C$^{18}$O of HBC 494. The envelope and the cavities are coloured with red to illustrated the red-shifted emission, while blue illustrates the blue shifted emission. Envelope material close to and accreting onto the disc is coloured with green and its infalling motion is indicated by the small green arrows. The green line with a position angle of $\sim$ 145$^{^{\circ}}$ depicts the rotation axis of the entire system.}
\label{Fig:art}
\end{figure*}

ALMA observations for HBC 494, located at 05$^{h}$ 40$^{s}$ 27.45$^{s}$ -07$^{o}$ 27$^{'}$ ${29.65}''$, were taken under program 2013.1.00710.S during Cycle-2 phase and over three different nights. This program involves the observation of eight FUor/EXor objects with results to be published in \citet{Cieza2016a}, \citet{Zurlo2016}, \citet{RuizRodriguez2016}, \citet{Principe2016} and \citet{Cieza2016}. The first two nights are December 12$^{th}$, 2014 and April 15$^{th}$, 2015 using 37 and 39 antennas on the C34-2/1 and C34/2 configurations, respectively. These configurations are quite similar with the shortest baseline of $\sim$ 14 m and longest of $\sim$ 350 m. The precipitable water vapor ranged from 0.7 to 1.7 mm with an integration time of $\sim$ 2 min per each epoch. Additionally, a third night, on August 30$^{th}$, 2015 HBC 494 was observed with 35 antennas in the C34-7/6 configuration with baselines ranging from 42 m to 1.5 km, an integration time of $\sim$3 min, and a precipitation water vapor of 1.2 mm. The quasars J0541-0541, J0532-0307 and/or J0529-0519 (nearby in the sky) were observed as phase calibrators.  J0423-013 and Ganymede were used as Flux calibrators, while the quasars J0607-0834  and J0538-4405  where observed for bandpass calibration.

Our correlator setup included the J=2-1 transitions of $^{12}$CO, $^{13}$CO and C$^{18}$O centered at 230.5380, 220.3987, and 219.5603 GHz, respectively. The correlator was configured to provide a spectral resolution of 0.04 km~s$^{-1}$ for $^{12}$CO and of 0.08 km~s$^{-1}$ for  $^{13}$CO and C$^{18}$O.  The  total bandwidth available for continuum observations was 3.9 GHz.  The observations from all three nights were concatenated and processed together to increase the signal to noise and $uv$-coverage. The visibility data were edited, calibrated and imaged in CASA v4.4 \citep{McMullin2007}. The uncertainty for calibrated flux is estimated to be 10 $\%$.  We used the CLEAN algorithm to image the data and using a robust parameter equal to zero, a briggs weighting was performed to adjust balance between resolution and sensitivity.  From the CLEAN process, we obtained the following synthesized beams:  ${0.35}''\times{0.27}''$with P.A. $=$ -90$^{^{\circ}}$ for $^{12}$CO, ${0.37}''\times{0.28}''$ with P.A. $=$ 86.5$^{^{\circ}}$ for $^{13}$CO and ${0.37}''\times{0.29}''$ with P.A. $=$ 87$^{^{\circ}}$ for C$^{18}$O. The rms is 12.5 mJy beam$^{-1}$ for $^{12}$CO, 16.0 mJy beam$^{-1}$ for $^{13}$CO and 13.9 mJy beam$^{-1}$ for C$^{18}$O. For the integrated continuum, we obtained a synthesized beam and rms of ${0.25}''\times{0.17}''$with P.A. $=$ -85.5$^{^{\circ}}$ and 0.25 mJy beam$^{-1}$, respectively. The maximum resolvable angle is 11 arc sec.

\section[]{Results}
\label{Sec:Results}

\begin{figure*}
\centering
\includegraphics[width=1.0\textwidth]{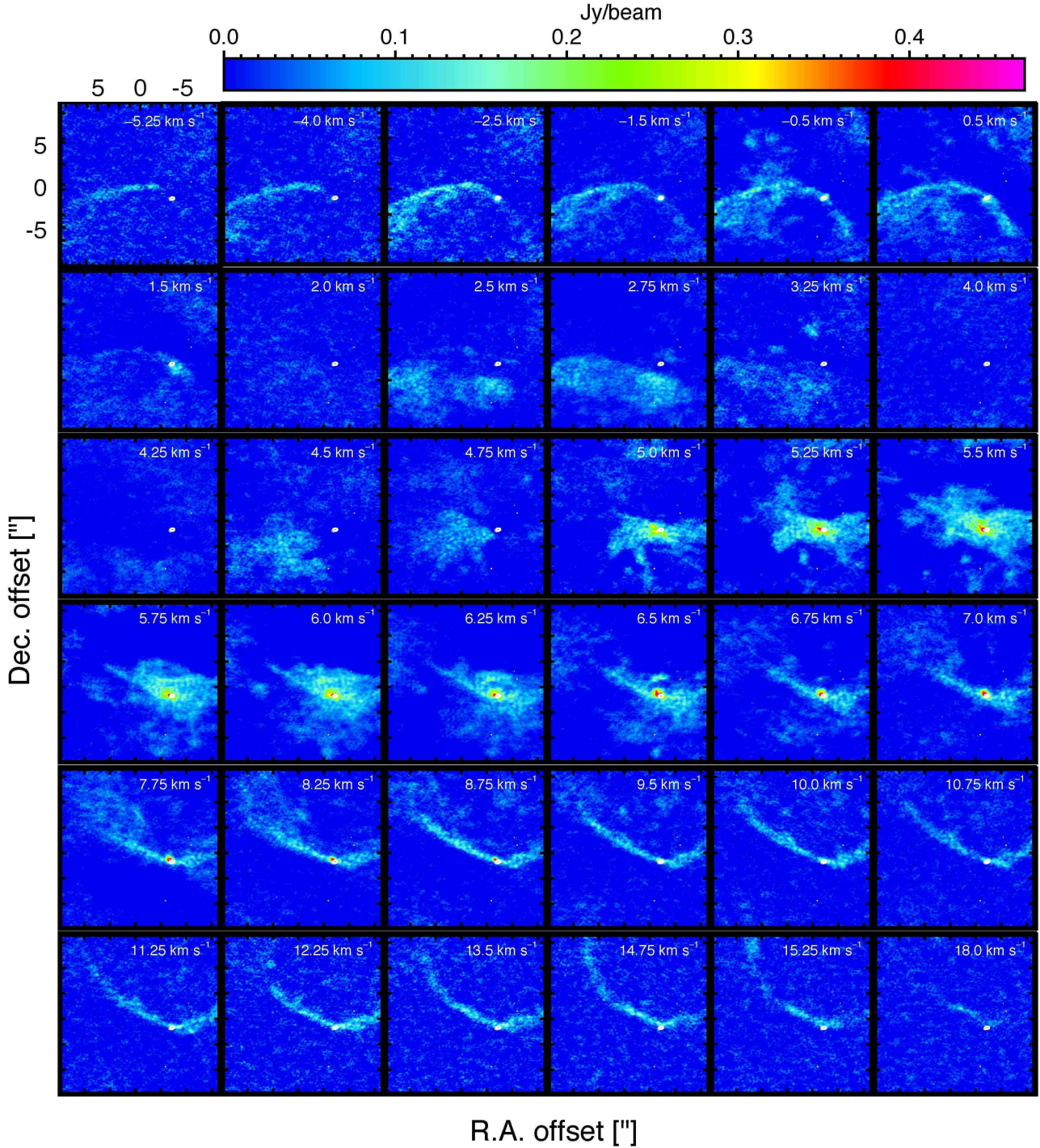}
\caption{Channel maps of the  $^{12}$CO. LSR velocities are shown at the top-right corner of each panel with a systemic velocity of $\sim$4.6 km~s$^{-1}$. White contours represent the  continuum emission around HBC 494 as shown in Figure \ref{Fig:CO12}.}
\label{Fig:CanalCO21}
\end{figure*}

Our data reveal the impressive extension of the outflow and envelope surrounding HBC 494 allowing us to piece together the main physical components of this object. The close edge-on configuration of the system means that the red- and blue-shifted outflow lobes are spatially separated. The complex gas kinematics and density gradients are traced by the blue-shifted and red-redshifted components of the CO emission. $^{12}$CO traces the bipolar and extension cavities of the outflow with a rotation axis oriented at $\sim$ 145$^{^{\circ}}$ \footnote{All position angles are specified north through east.} (Section \ref{Sec:co12results}). $^{13}$CO probes the infalling and rotating envelope surrounding the protostar and disc. A fraction of the mass of this part of the envelope is eventually transported onto the disc to be accreted onto the protostar, while at more distant regions from the central object, another fraction is being pushed away to external regions by the opposing outflow (Section \ref{Sec:co13results}). C$^{18}$O traces similar regions as $^{13}$CO at the northern side of the central object, while at larger distances, the envelope  is driven out by the wider-angle portions of the wind (Section \ref{Sec:co18results}). Also, the C$^{18}$O is weak at the southern region of the system, suggesting lower densities. Using the C$^{18}$O line, we estimated a systemic velocity of $\sim$4.6 km~s$^{-1}$, see section \ref{sec:lines}. We further estimate the kinematics and masses of the outflow and envelope from the emission and its velocity structure (Section \ref{Sec:kinematics}). To picture HBC 494 in a more comprehensive way, we provide a cartoon showing the main components drawn from the $^{12}$CO, $^{13}$CO and C$^{18}$O emissions, see Figure \ref{Fig:art}.

\subsection{$^{12}$CO Moment Maps}
\label{Sec:co12results}

\begin{figure*}
    \centering
    \begin{subfigure}[b]{0.495\textwidth}
        \centering
                 \includegraphics[width=1.0\textwidth]{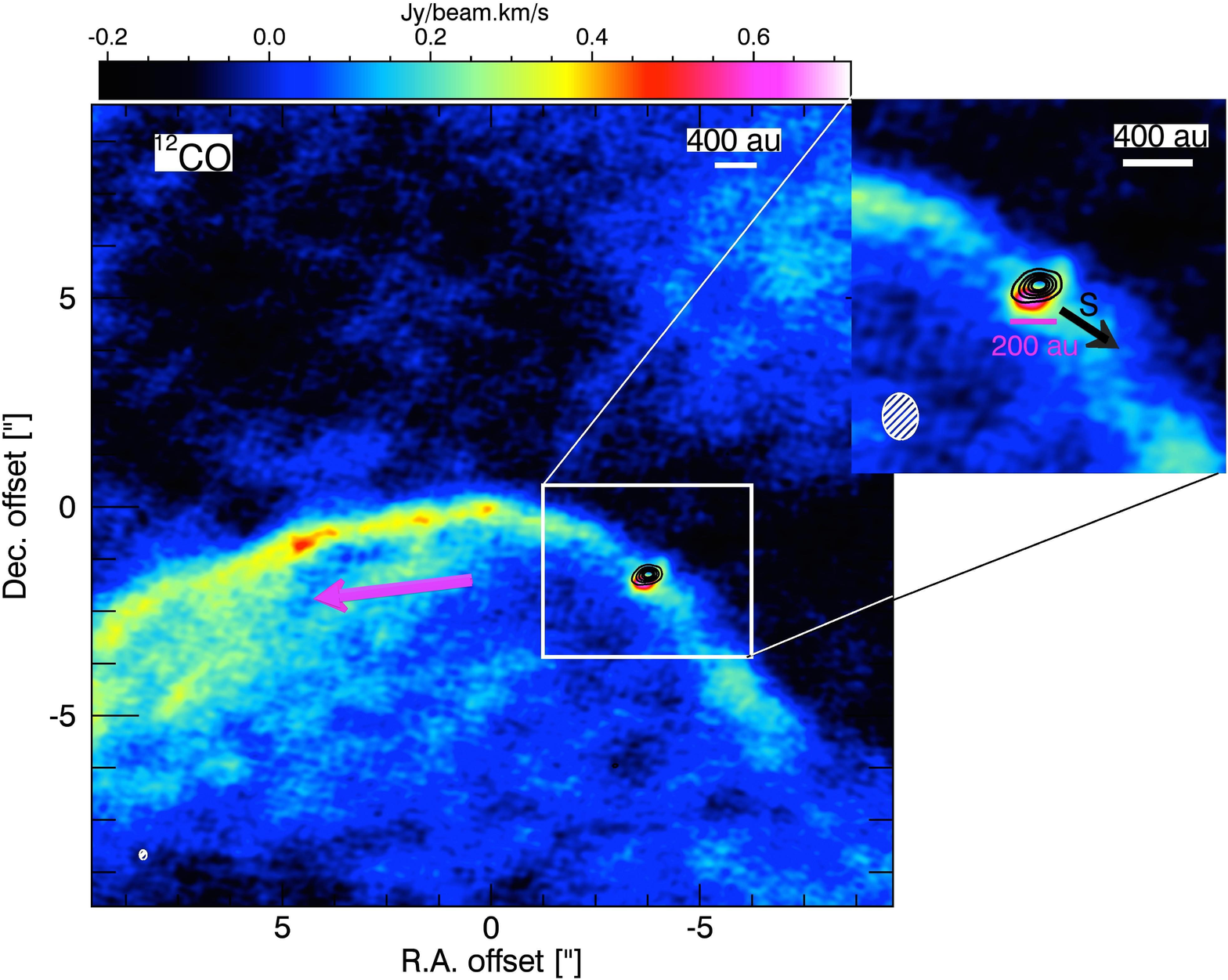}
                 \caption{Moment 0: Southern Outflow}
        \label{Fig:CO12a}
    \end{subfigure}
    \hfill
    \begin{subfigure}[b]{0.495\textwidth}
        \centering
        \includegraphics[width=1.0\textwidth]{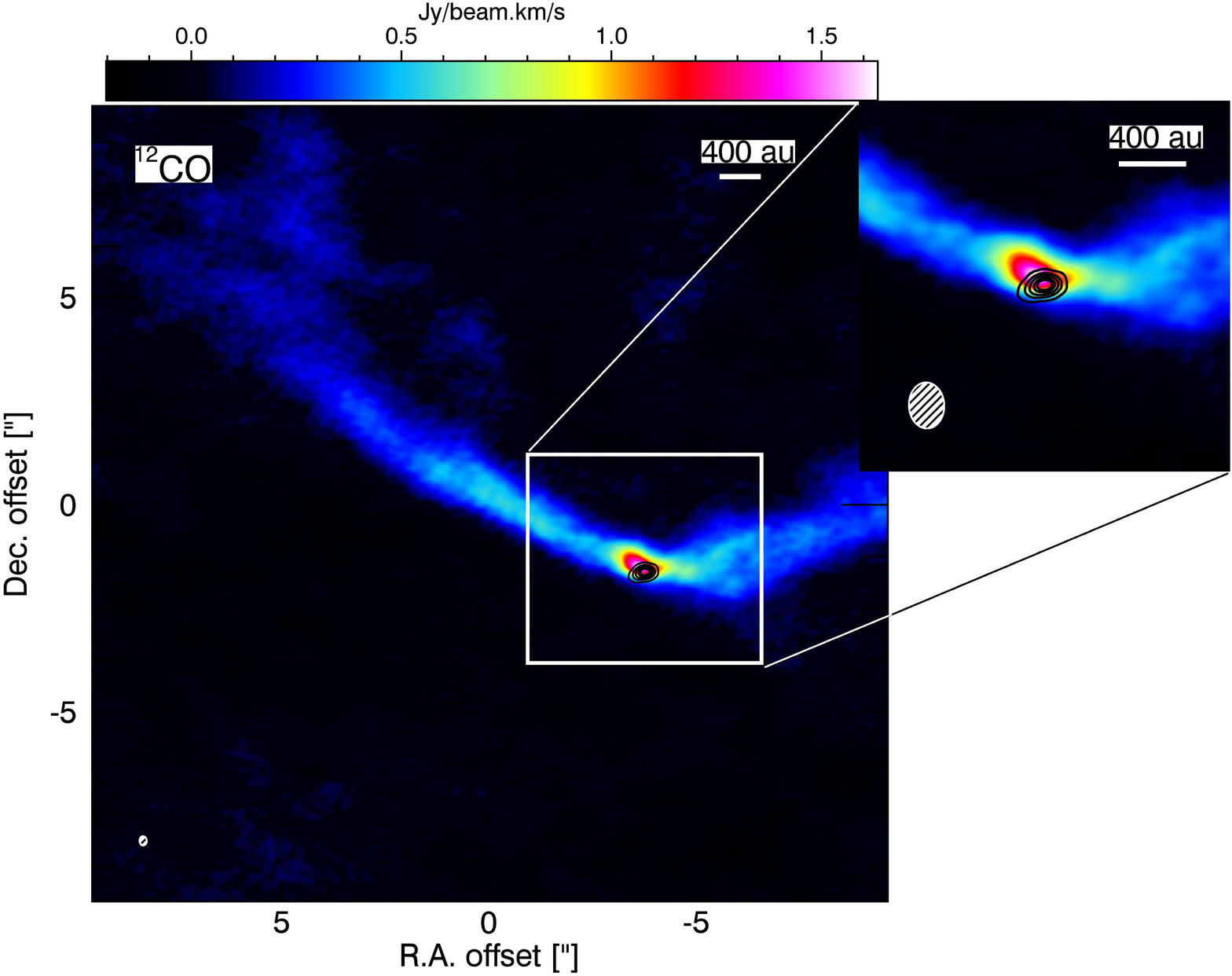}
         \caption{Moment 0: Northern Outflow}
        \label{Fig:CO12b}
    \end{subfigure}
    \hfill
    \begin{subfigure}[b]{0.6\textwidth}
        \centering
                \includegraphics[width=0.9\textwidth]{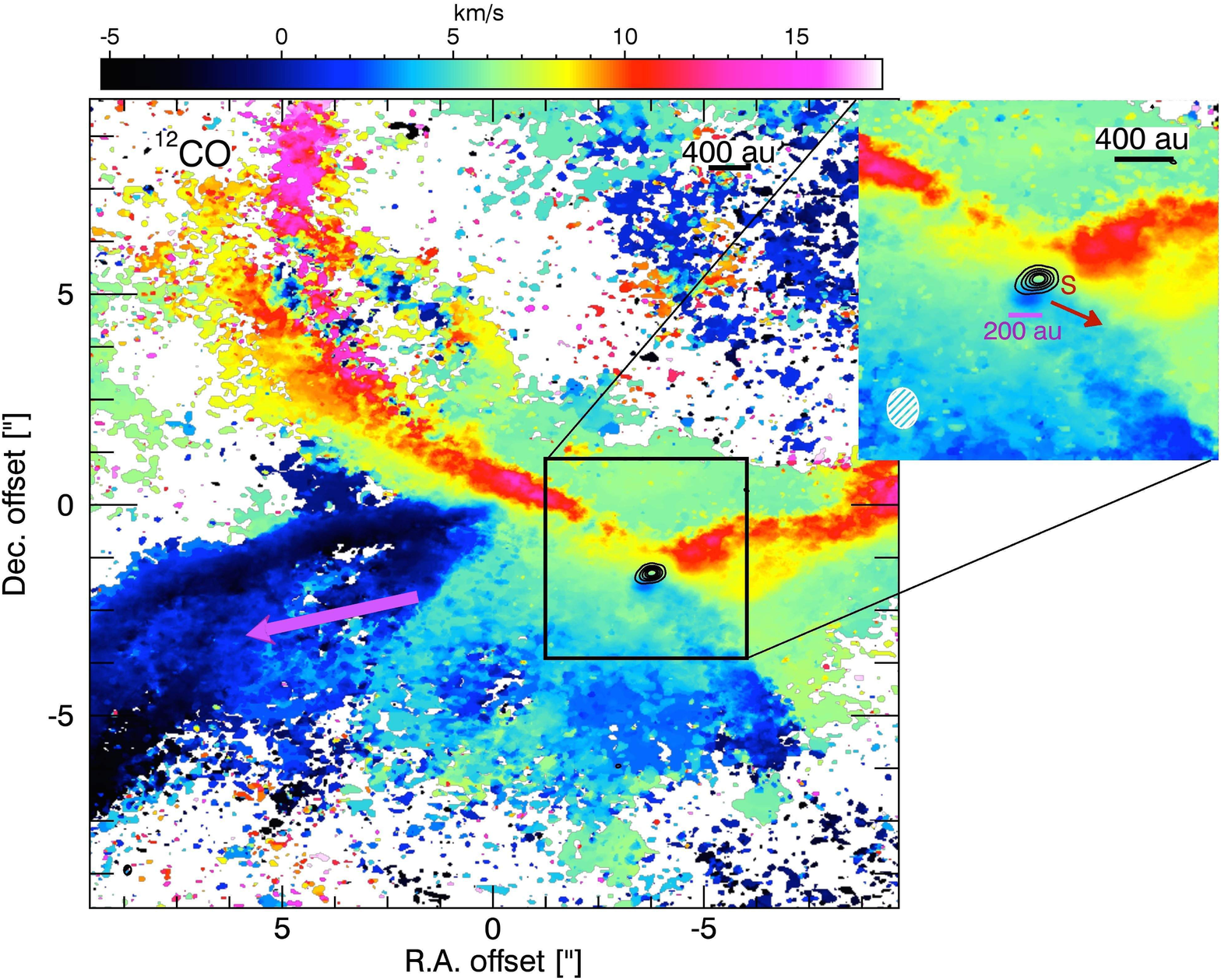}
                \caption{Moment 1: Entire System. }
        \label{Fig:CO12c}
    \end{subfigure}    
       \caption{Figure a: Integrated intensity maps of the $^{12}$CO blue- shifted emission from -5.25 to 4.75 km~s$^{-1}$ channels. Figure b: Integrated intensity maps of the $^{12}$CO red- shifted emission on the velocity range between 7.0 and 18 km~s$^{-1}$ . Figure c: $^{12}$CO velocity field map that was obtained from the integration over the velocity range from -5.5 to 18 km~s$^{-1}$. Black contours show the continuum emission around HBC 494 at 10, 30, 80, 150 and 250 $\times$ rms (0.25 mJy beam$^{-1}$). The ${0.35}''\times{0.27}''$with P.A. $=$ -90$^{^{\circ}}$ synthesised beam is shown on the lower left corner of each panel. The upper right insets are a closeup ($\pm$ {2.7}'') of the central object. The black and red arrows shown in the insets of Figures  \ref{Fig:CO12a} and \ref{Fig:CO12c} point out the ``stream'' described in section \ref{Sec:co12results}. While the purple arrow shows the material interacting with the surrounding envelope, also detected at $^{13}$CO and C$^{18}$O  emissions, see Figures \ref{Fig:CO13} and \ref{Fig:CO18}.} 
    \label{fig:three graphs}
  \label{Fig:CO12}
\end{figure*}

The $^{12}$CO emission traces the highly energetic outflow blowing through the gas and creating a bipolar cavity in the molecular cloud. In our data cubes, significant $^{12}$CO emission is detected at the velocities ranging from -5.25 to 18 km~s$^{-1}$, see Figure \ref{Fig:CanalCO21}. We integrated separately, the channels corresponding to the ``Northern'' and the ``Southern'' outflows to show structural shapes in a more clear manner. The moments 0 and 1 of the high-velocity blue- and red-shifted outflow  cavities are shown in Figures \ref{Fig:CO12a}, \ref{Fig:CO12b} and \ref{Fig:CO12c} and their corresponding zoomed version in the small windows next to each figure. Figure \ref{Fig:CO12a} and \ref{Fig:CO12b} have a velocity range, with respect to the Local Standard of Rest (LSR), between -5.25 and 4.75 km~s$^{-1}$ for the blue-shifted region and between 6.75 and 18 km~s$^{-1}$ for the red-shifted emission. Although the complete extension of these bipolar cavities is not observed because it falls outside of our field of view, the very \textit{wide} outflow cavities reach an apparent opening angle of $\sim 150$$^{^{\circ}}$with an extension of at least 8000 au at a distance of 415 pc. Such a sculpted and defined outflow cavity is among the widest known to date and is remarkable for a Class I object, which typically have narrower cavities at these early stages. The overall range of these Class I outflows varies between 30 and 125$^{^{\circ}}$  \citep{Arce2006, Klaassen2016, Zurlo2016, Principe2016}. {In addition, the $^{12}$CO emission comes predominantly from the material directly influenced by the outflow, indicating the deep sweeping of surrounding material inside out acquiring a higher temperature. This might be a product of the interaction between the surrounding material and the high luminosity central protostar \citep[$\sim$ 250 $\Lsun$;][]{Reipurth1986}. The lack of uniform emission in the extension of the cavities as seen in Figures \ref{Fig:CO12a} and  \ref{Fig:CO12b}, is likely related to the ALMA maximum recoverable scale of 11$\rm ''$ that would correspond to $\sim$ 4500 au and prevents resolving larger scale structures. Thus, the ``missing'' emission does not imply a lack of $^{12}$CO gas emission in between cavity arms \citep[e.g.][]{Bradshaw2015}. 

\begin{figure*}
\centering
\includegraphics[width=0.7\textwidth]{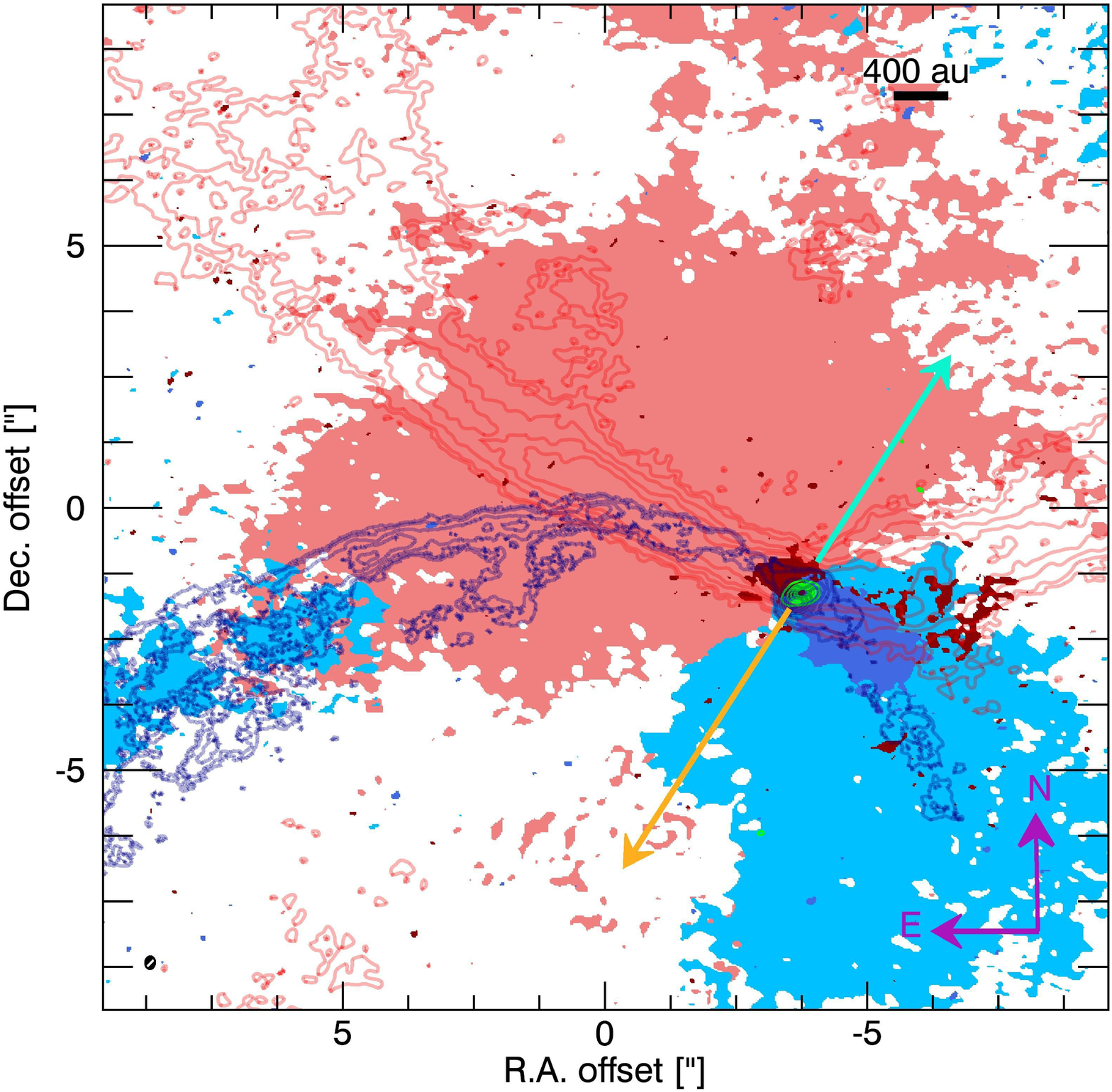}
\caption{Comparison of the  $^{12}$CO and $^{13}$CO velocity range-integrated intensity maps (moment 0). Blue and red contours show the integrated intensity of the $^{12}$CO blue- and red shifted lobes, respectively, at 80, 150, 300, 450 and 600 $\times$ 3$\sigma$ levels. Light and dark red regions correspond to the red-shifted low (extended emission) and high (compact emission) velocities, respectively. Light and dark blue regions correspond to the blue-shifted low (extended emission) and high (compact emission) velocities, respectively. The extended and compact emission covers in blue-shifted emission velocities between 1.0 and 1.75 km/s and 2.5 and 3.5 and in red-shifted emission between 6.5 and 8.0 km/s and 4.5 and 5.5, which are explained in Section \ref{Sec:co13results}. Green contours show the continuum emission around HBC 494 at 10, 30, 80, 150 and 250 $\times$ rms (0.25 mJy beam$^{-1}$). The synthesized beam is shown on the lower left corner. Cyan and yellow arrows show the projected axis alignment along the Southern outflow - disc - Northern outflow system.
}
\label{Fig:CO13grey}
\end{figure*}

 The red-shifted velocity field of $^{12}$CO, considering emission merely from the outflow, can be described in terms of the degree of interaction with the surrounding envelope. The Northern outflow presents a velocity pattern with a gradient field perpendicular to the outflow axis that ranges between 7.0 and 18 km~s$^{-1}$. This suggests that its opening angle increases as the abundant envelope material is being removed inside out from the outflow axis. Additionally, at the base of the bipolar cavities, a large amount of material with velocities from 5.0 to 6.75 km~s$^{-1}$ seems to follow an infalling motion surrounding the central object, see Figure \ref{Fig:CanalCO21} and Figure \ref{Fig:CO12c}. This red-shifted outflow emission at velocities close to the ambient molecular cloud is dense gas being impacted by the outflow that belongs to the infalling envelope, which might be composed of ionized and hot material as a result of the direct interaction with the launched outflows. This slab of material closer to the rotation plane partially overlaps with the $^{13}$CO emission tracing medium densities corresponding to the envelope material, a more detailed explanation can be found in Section \ref{Sec:co13results}.

The $^{12}$CO blue-shifted emission probes at the Southern side of HBC 494 a complex geometry of the outflowing molecular material. The emission arises from the accelerated molecular gas depending on the medium properties (e.g. geometry, density). Indeed, it is detected at a velocity range between -5.25 and 1.5 km~s$^{-1}$ that the Southern cavity shows a brighter region, more concentrated at the southwest of the object, possibly due to the interaction with a significant amount of gas in the surrounding envelope, see purple arrow in Figure \ref{Fig:CO12}. While, the $^{12}$CO blue-shifted emission between 2.5 and 3.5 km~s$^{-1}$ might be outflow-envelope interactions with expanding motions.

In contrast to the wide angle outflow, it can be noted in the moments 0 and 1 maps and shown in the insets of Figures \ref{Fig:CO12a} and \ref{Fig:CO12c},  an ``intensity maximum'' with a diameter of around $\sim$ 200 au that lies at the southwest from the continuum emission. This intensity maximum has a blue-shifted component that reaches 2.9 km~s$^{-1}$ and coincides with the location of the asymmetry found in the continuum by \citet{Cieza2016}, see Figure \ref{Fig:continuum}. However, it is not straightforward to attribute a physical origin to this feature and future observations with higher resolution are required.

The integrated flux on both sides of the bipolar outflow differ by a factor of $\sim$2.5. At the Northern cavity the integrated flux is 39.65 $\pm$ 0.09 Jy km~s$^{-1}$ and at the Southern cavity a value of 93.32 $\pm$ 0.12 Jy km~s$^{-1}$, see Figures \ref{Fig:CO12a} and Figure \ref{Fig:CO12b}. This might be evidence of the non-uniformity of the molecular cloud, where the evolution of this Class I object is taking place. Considering that CO transitions are thermalised at or close to their critical densities \citep[$\sim$ 1.1 x 10$^{4}$ cm $^{-3}$;][]{Carilli2013}, the absence of a stronger emission at the Northern region indicates a slightly denser cloud material; while the stronger emission at the Southern region might point out a more widespread region that interacts with the ejected material from the central object. Additionally, the alignment among the South outflow $-$ disc $-$ North outflow at the base of both lobes, is evident in their $^{12}$CO contours delineating the limb-brightened walls of the parabolic outflow cavities and the continuum, see Figure \ref{Fig:CO13grey}. This alignment allows us to draw a line through it and then, compute the outflow position angle (PA) of $\sim$ 145$^{^{\circ}}$ north through east.

\subsection{$^{13}$CO  Moment Maps}
\label{Sec:co13results}

\begin{figure*}
\centering
\includegraphics[width=1.0\textwidth]{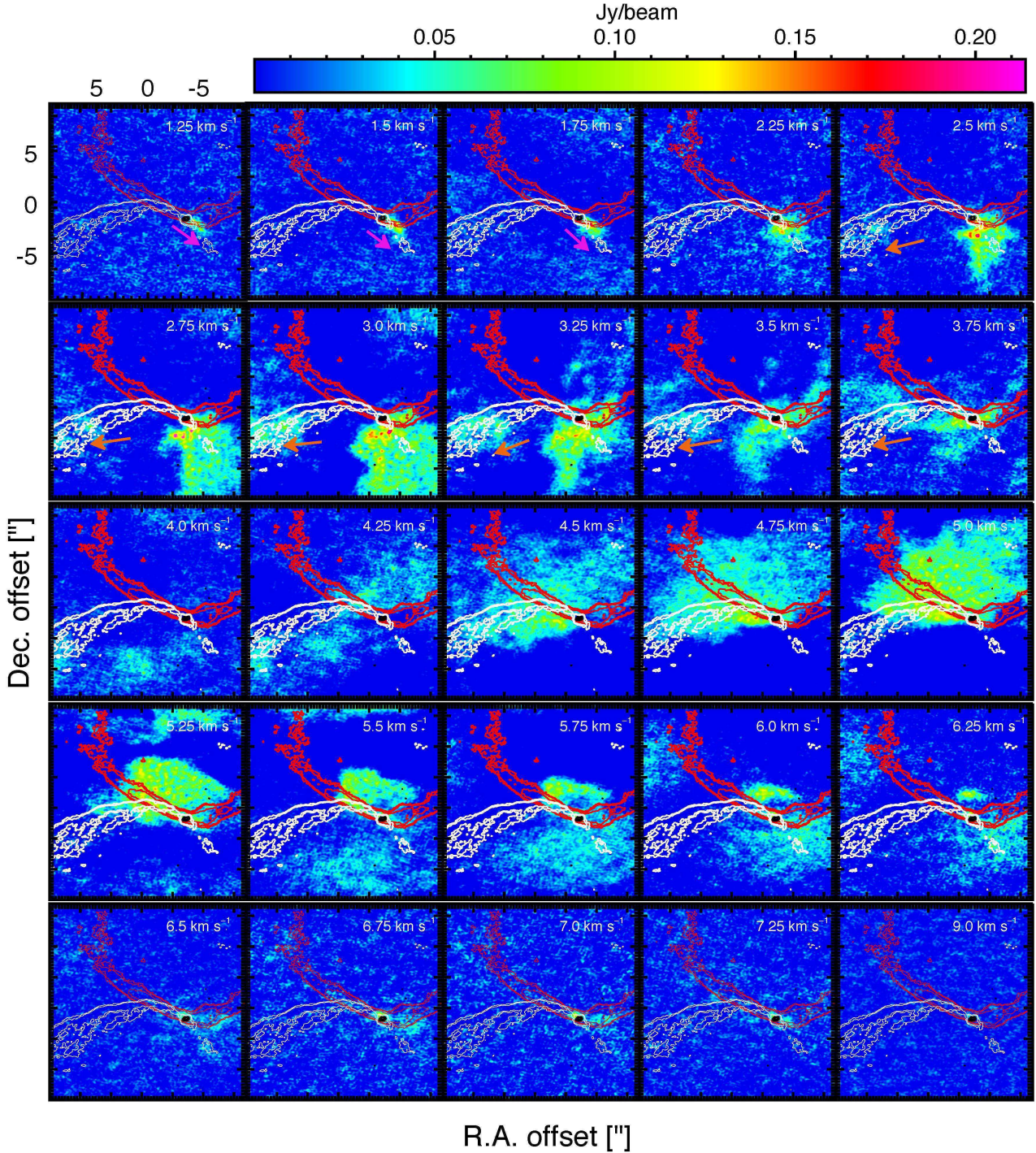}
\caption{Channel maps of the  $^{13}$CO. LSR velocities are shown at the top-right corner of each panel with a systemic velocity of $\sim$4.6 km~s $^{-1}$. White and red contours show the integrated intensity of the $^{12}$CO blue- and red shifted lobes, respectively, at 150 and 300 $\times$ 3$\sigma$ levels. Magenta arrows in the channels with velocities from 1.25 to 1.75 km~s $^{-1}$ point out the emission of the streamline on this side of the system. Brown arrows in the channels with velocities from 2.5 to 3.75 km~s $^{-1}$ point out the interaction between the Southern outflow and the surrounding envelope and showing how as a result there is expelling material on this side of the envelope directed toward us. Black contours show the continuum emission around HBC 494 at 10, 30, 80, 150 and 250 $\times$ rms.}
\label{Fig:CanalCO13}
\end{figure*}

\begin{figure*}
    \centering
    \begin{subfigure}{0.495\textwidth}
        \centering
        \includegraphics[width=1.0\textwidth]{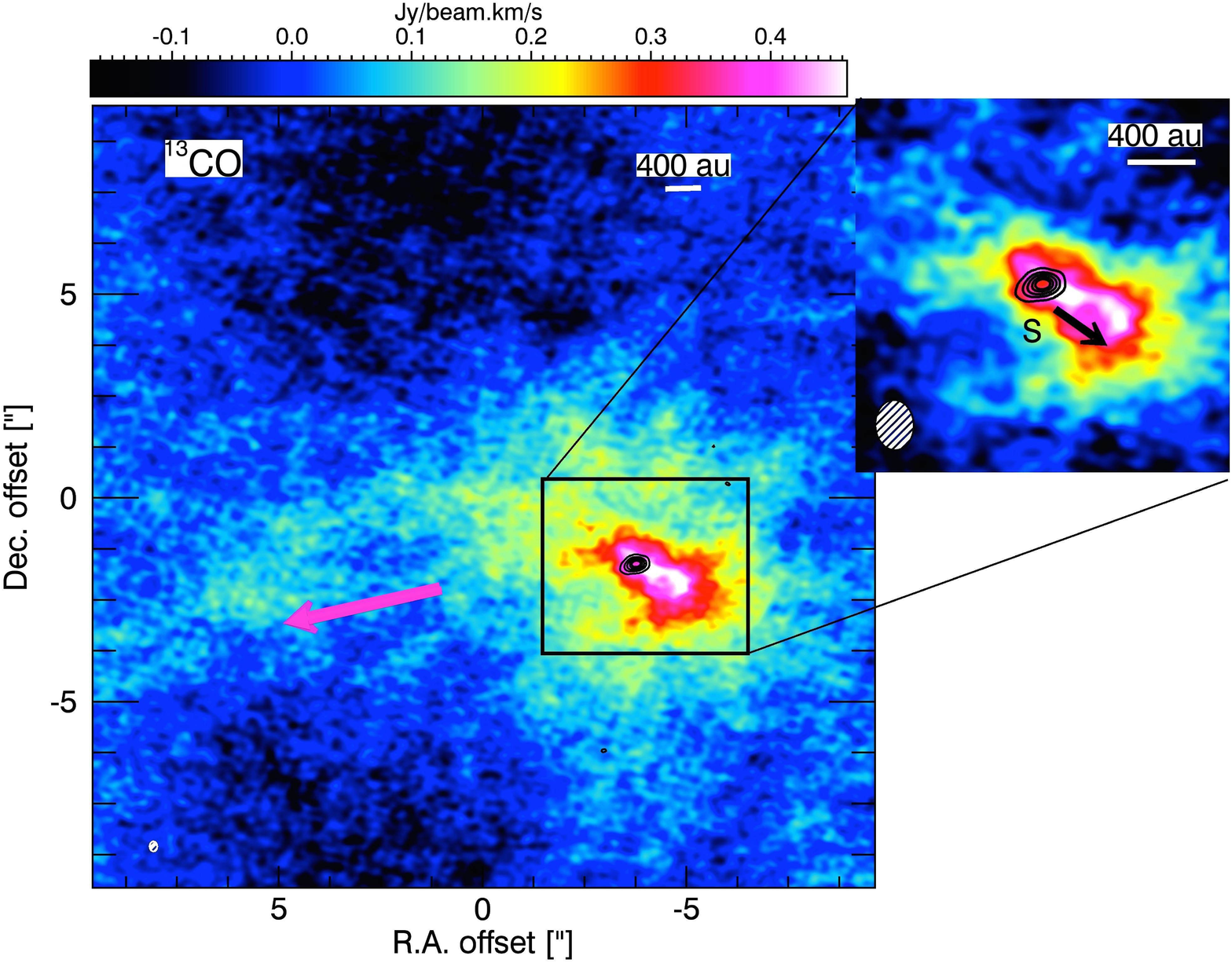}
        \caption{Moment 0}
        \label{Fig:CO13a}
    \end{subfigure}
      \hfill
    \begin{subfigure}{0.495\textwidth}
        \centering
        \includegraphics[width=1.0\textwidth]{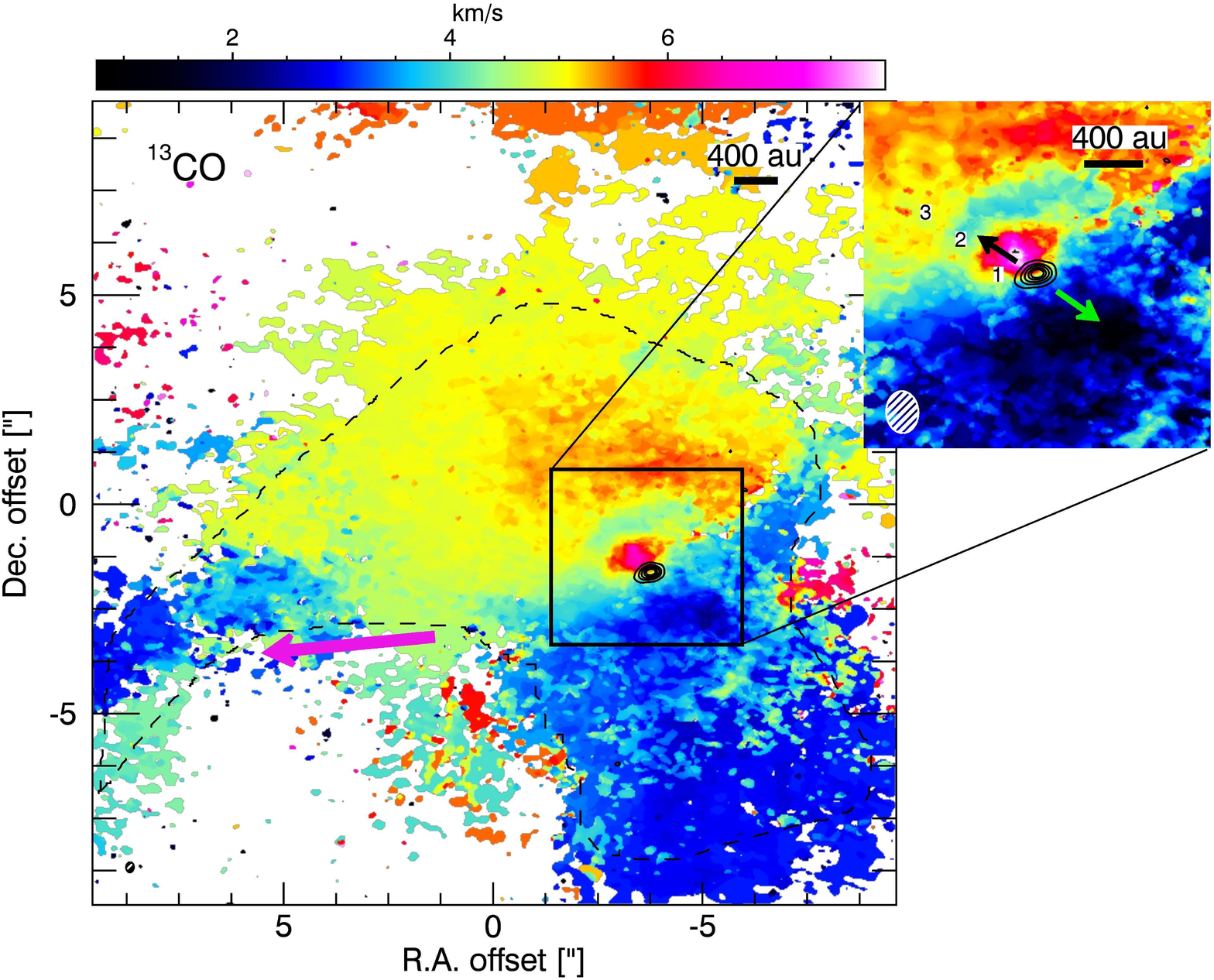}
        \caption{Moment 1}
        \label{Fig:CO13b}
    \end{subfigure}
       \caption{Figure a: $^{13}$CO  intensity maps (moment-0) integrated over the velocity range of 1.0 $-$ 9.0 km~s$^{-1}$. Figure b: Intensity - weighted mean velocity  (moment-1) map of $^{13}$CO  in the velocity range 1.0 $-$ 9.0 km~s$^{-1}$ over values higher than a 3$\sigma$ ($\sigma \sim$ 15 mJy beam$^{-1}$ km s$^{-1}$). The ${0.37}''\times{0.28}''$ with P.A. $=$ 86.5$^{^{\circ}}$ synthesised beam is shown on the lower left corner of each panel. Black contours show the continuum emission around HBC 494 at 10, 30, 80, 150 and 250 $\times$ rms (0.25 mJy beam$^{-1}$). The upper right insets are a closeup ($\pm$ {2.7}'') of the central object. The black and green arrows shown in the insets of Figures  \ref{Fig:CO13a} and \ref{Fig:CO13b} point out the streamline described in section \ref{Sec:co13results}. While the purple arrow shows the material interacting with the outflow detected at $^{12}$CO emission, see Figures \ref{Fig:CO12}. The region enclosed with dashed lines correspond to the region in which we integrated the line profile shown in Figure \ref{Fig:spectra}.}
    \label{Fig:CO13}
\end{figure*}

We find that the $^{13}$CO emission traces the rotating, infalling and expanding envelope surrounding the central system. The blue-shifted emission is detected at velocities from 1.0 to 4.25 km~s$^{-1}$ and the red-shifted material has a range of velocities from 4.5 to 9.0 km~s$^{-1}$ (Figure \ref{Fig:CanalCO13}). Figure \ref{Fig:CO13a} and \ref{Fig:CO13b} shows the moment-0  and  moment-1 maps of the $^{13}$CO line, while the small windows are a zoomed image to the central object. Since the $^{13}$CO shows a complex structure with compact and extended emission that comes from different regions of the envelope, we integrated channel maps of blue- and red-shifted low (extended emission) and high (compact emission) velocities in order to see a more detailed velocity structure and a direction of the velocity gradient \citep[e.g.][]{Aso2015}. Figure \ref{Fig:CO13grey} shows the moment 0 with integration of the blue-shifted emission at velocities between 1.0 and 1.75 km s$^{-1}$ and 2.5 and 3.5 km s$^{-1}$ and red-shifted emission between 6.5 and 8.0 km s$^{-1}$ and 4.5 and 5.5 km s$^{-1}$. The light red and blue and dark blue and red areas correspond to the extended and compact regions, respectively. Using the compact emission, we draw a line along the gradient velocity and then, estimate a PA of $\sim$ 50$^{^{\circ}}$ north through east, which is almost perpendicular to the outflow PA of $\sim$ 145$^{^{\circ}}$. Similarly, the low velocity components (extended structure) show a velocity gradient with a PA of $\sim$ 50$^{^{\circ}}$.

The compact blue-shifted emission protrudes with a high flux density that we named as ``stream'' (S) located at the southeast of the disc and indicated by the black arrow along the feature in Figure \ref{Fig:CO13a}. In Figure \ref{Fig:CO13b} a green arrow is shown representing the location of the stream detected in the moment 0. The direction of this compact structure is difficult to indicate, but it might have originated in wide-angle winds from the central object blowing into the envelope \citep{Snell1980, Shu2000, Gardiner2003} and expelling material to larger radii, following the Southern outflow. Figure \ref{Fig:CanalCO13} presents the channel maps of the $^{13}$CO line, where this stream or compact blue-shifted emission is indicated with a magenta arrow and $^{12}$CO contours are over plotted to spatially compare these emissions. On the other hand, the extended blue-shifted emission is more likely to be part of the surrounding envelope, being pushed toward us by the outflow. In addition, on the bottom-left side of Figure \ref{Fig:CO13b} an elongated structure appears, indicated with a purple arrow, and with blue-shifted velocities of 2.5 and 3.75 km~s$^{-1}$, coincides with the $^{12}$CO emission detected in the Southern region of a velocity range of 2.5 $-$ 3.25 km~s$^{-1}$. Thus, the ``extended'' emission is likely tracing the internal structure of the envelope and showing how the southern outflow is expelling material on this side of the envelope directed toward us.

The $^{13}$CO red-shifted emission also originates from two different regions in the envelope: one close and other farther away from the central system. The red-shifted compact emission source is the envelope material located very close to the protostar that reaches velocities of 6.5 and up to 9 km~s$^{-1}$ (Figure \ref{Fig:CanalCO13}). The origin of this emission is likely related to the dragged gas by the ejection of matter at the Northern cavity and indicated in Figure \ref{Fig:CO13b} as region 1. The second region, shown as regions 2 and 3 in the inset of Figure  \ref{Fig:CO13b}, suggests that the emission at velocities between $\sim$4.5 and 6.25 km~s$^{-1}$ corresponds to material being accreted onto the system and thus, indicates the kinematics of the infalling and rotating envelope at the base of the bipolar outflows. As mentioned in Section \ref{Sec:co12results}, the slab of material closer to the rotation plane partially overlaps with the $^{12}$CO emission at the base of the cavities, thus strongly suggesting their accreting nature onto the central object. In addition, the emission in the velocity range between $\sim$5.5 and 6.25 km~s$^{-1}$ extends around the central system forming a feature with the shape of a half symmetric ``ring'' with a distance between the inner and outer ring of around $\sim$ 200 au, shown as region 2 in Figure \ref{Fig:CO13b}. Although, the origin of this feature is not clear, we speculate that it is a signature of material being accreted with a rotation motion.

\subsubsection{Position-Velocity Diagram}

The velocity gradient in the emission of $^{13}$CO  is perpendicular to the direction of the outflow. Thus, we have created a position$-$velocity (PV) diagram by cutting along the axis perpendicular to the rotating outflow and throughout the continuum emission of the disc, see Figure \ref{Fig:pv}. In Figure \ref{Fig:CO13grey} the cyan+yellow line indicates the outflow rotation axis and perpendicular to it, the image-space PV diagram cut at a position angle of $\sim$ 50$^{^{\circ}}$ and the averaging width is $\sim$ 0.45$^{''}$. This diagram shows higher velocities toward the center of the system as expected for envelope material feeding a central proto-stellar source.  

\begin{figure}
\centering
\includegraphics[width=0.47\textwidth]{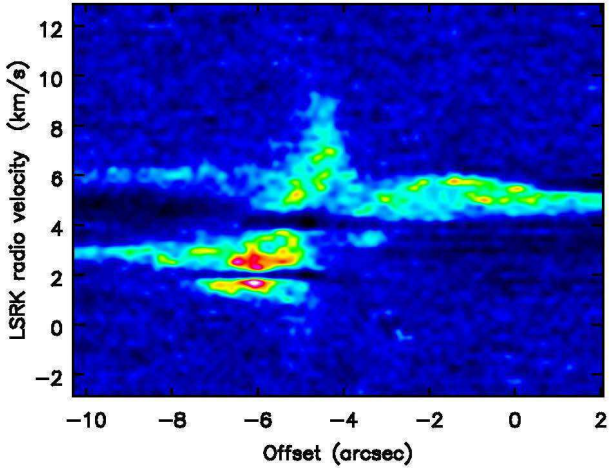}
\caption{Position - velocity map of $^{13}$CO along the axis perpendicular to the rotating outflow, see Figure \ref{Fig:CO13grey}.
}
\label{Fig:pv}
\end{figure}

\subsection{C$^{18}$O  Moment Maps}
\label{Sec:co18results}

\begin{figure}
    \centering
    \begin{subfigure}[b]{0.495\textwidth}
        \centering
        \includegraphics[width=1.0\textwidth]{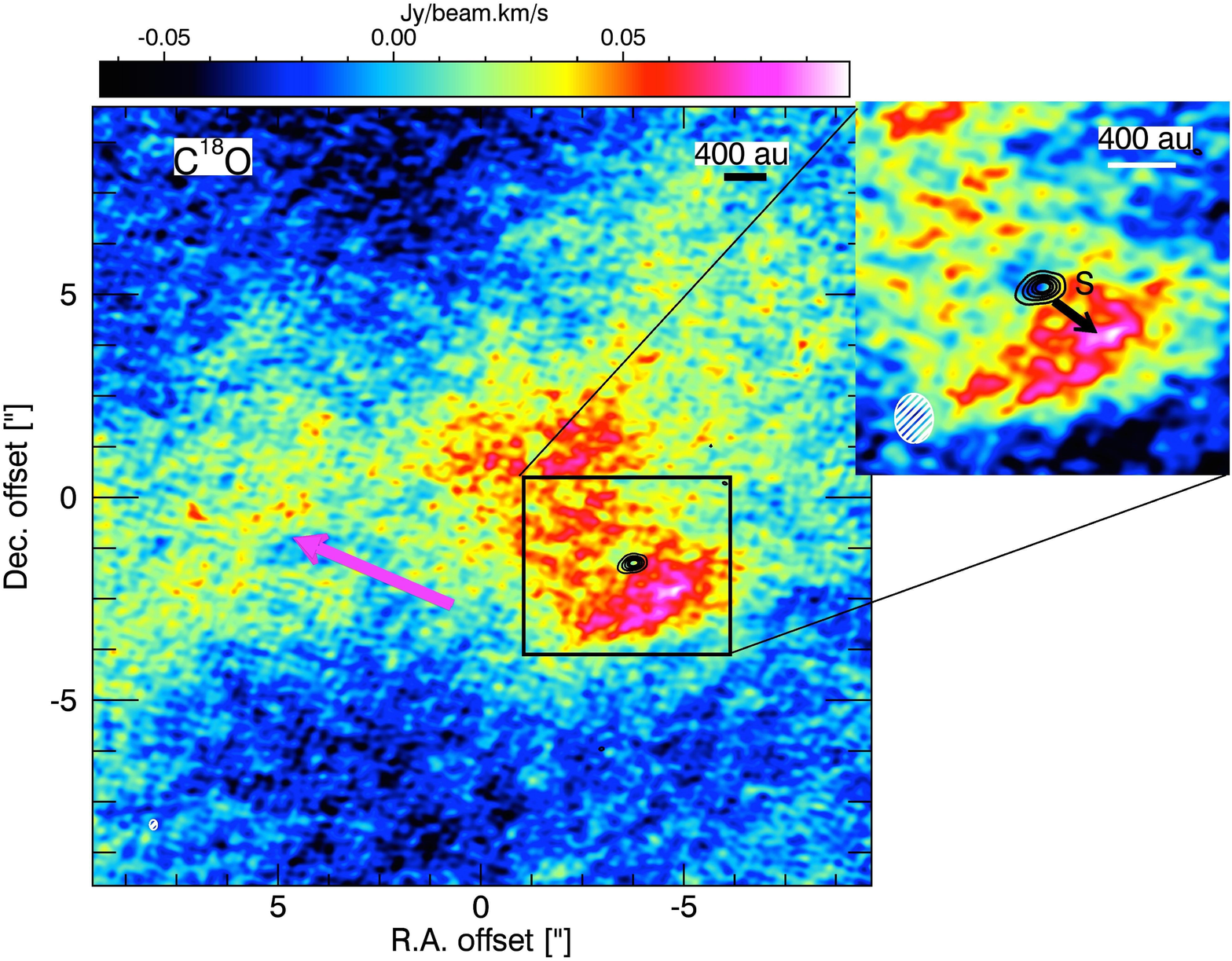}
        \caption{Moment 0}
        \label{Fig:CO18a}
    \end{subfigure}
    \hfill
    \begin{subfigure}[b]{0.495\textwidth}
        \centering
        \includegraphics[width=1.0\textwidth]{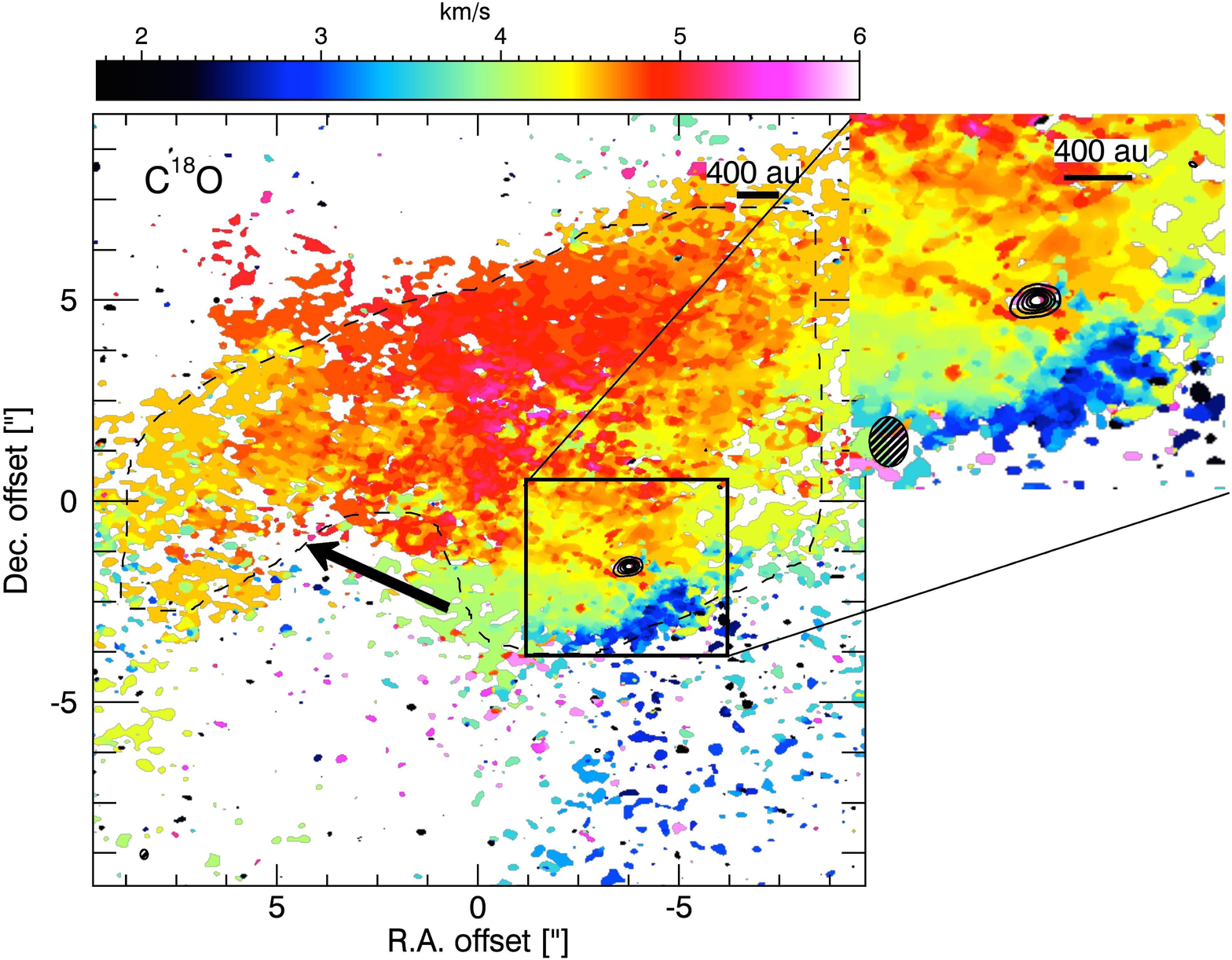}
        \caption{Moment 1}
        \label{Fig:CO18b}
    \end{subfigure}
   
    \caption{Figure a: Integrated C$^{18}$O  intensity over the velocity range between 1.75 and 6 km~s$^{-1}$. Figure b: Intensity weighted mean velocity. The ${0.37}''\times{0.29}''$ with P.A. $=$ 87$^{^{\circ}}$ synthesized beam is shown on the lower left corner of each panel. Black contours show the CO continuum emission around HBC 494 at 10, 30, 80, 150 and 250 $\times$ rms (0.25 mJy beam$^{-1}$). The upper right insets are a closeup ($\pm {2.7}''$) of the central object. The black arrow shown in the inset of Figure \ref{Fig:CO18a} point out the streamline described in section \ref{Sec:co18results}. While the purple and black arrows show the material interacting with the outflow detected at $^{12}$CO emission, see Figures \ref{Fig:CO12}. The region enclosed with dashed lines correspond to the region in which we integrated the line profile shown in Figure \ref{Fig:spectra}.}
    \label{Fig:CO18}
\end{figure}

Of the 3 isotopologues, C$^{18}$O  has the lowest abundances and thus traces higher gas density regions inside the molecular cloud, where it sublimates off dust grains.  Therefore, we used this line to map the morphology of the envelope surrounding HBC~494. With a channel width of 0.25 km~s$^{-1}$, the emission is detected within a range of 1.75 and 6 km~s$^{-1}$. Figure \ref{Fig:CO18a} shows the integrated flux over the spectral line and Figure \ref{Fig:CO18b} is the intensity-weighted velocity of the spectral line. The blue-shifted outflow is seen at velocities from 1.75 to 4.25 km~s$^{-1}$. The stream feature detected in $^{13}$CO  at 1$-$1.75 km~s$^{-1}$ is also seen in C$^{18}$O emission at 1.75-2.0 km~s$^{-1}$, see Figure \ref{Fig:CO18a} for stream location, while the C$^{18}$O emission in the range of 2.25 and 4.25 km~s$^{-1}$ traces the shape of the envelope. Additionally, the C$^{18}$O blue-shifted emission is very weak compared to the red-shifted emission and is located mostly at regions close to the central source. The faint or lack of blue-shifted emission at larger distances from the central object in the southern direction might be an indication of a significant extended and diffuse gas, where the emission origin corresponds to a break out of the surrounding molecular cloud. 

The C$^{18}$O red-shifted emission detected at velocities from 4.5 and 6 km~s$^{-1}$ is bright and overlaps the $^{13}$CO red-shifted emission, see Figure \ref{Fig:CO18b}, meaning that this probes infalling and expanding envelope material at different locations from the central source. Following our assumption that the northern side of the cavity is more denser than the southern side, given the lower $^{12}$CO emission at the northern outflow, then this traces out the colder and denser structures composing the infalling envelope. In addition, gas swept up by the southern outflow is seen to the west with low velocities of around $\sim$ 4.25 and 4.75 km~s$^{-1}$ and connected to rotating material in between the limbs of the outflows. The swept up gas is located slightly above the $^{13}$CO arm emission and indicated with purple and black arrows in Figures  \ref{Fig:CO18a} and \ref{Fig:CO18b}.

\subsection{Outflow Masses and  Kinematics}
\label{Sec:kinematics}

We use our molecular line data to derive the mass and kinematics of the outflow. However, as demonstrated in \citet{Cabrit1990}, the estimated outflow masses and dynamical properties computed from an optically thick line such as  $^{12}$CO can be considerably underestimated if they are not corrected for optical depth effects ($\tau_{12}$). Therefore, before computing the outflow properties, we corrected our molecular line data following standard methods such as those found in \citet{Arce2001}, \citet{Curtis2010} and \citet{Dunham2014} to evaluate $\tau_{12}$ numerically. Essentially, these methods are based on computing abundance ratios of optically thin CO emissions. Then, assuming identical beam-filling factors and the same excitation temperature for both isotopes, also considering that $^{13}$CO traces the optically thin emission of the outflow in detections at low velocities, we compute the ratio of the brightness temperatures between $^{12}$CO and $^{13}$CO, $\frac{\textrm{T}_{12}}{\textrm{T}_{13}} = X_{12,13}\frac{1- \textrm{exp}(-\tau _{12})}{\tau _{12}}$, where the abundance ratio $X_{12,13}$ = [$^{12}$CO]/[$^{13}$CO] is taken as 62 \citep{Langer1993}. $\frac{\textrm{T}_{12}}{\textrm{T}_{13}}$ in each channel is estimated by computing the weighted mean values, where the weight was performed using the sigma values of every channel. To compute the $^{12}$CO mass, we apply the correction factor to all the channels with $^{13}$CO detection above 5$\sigma$. For those channels where $^{13}$CO is too weak to be detected, we extrapolate values from a parabola fitted to the weighted mean values of the form:

\begin{displaymath}
\centering
\frac{\textrm{T}_{12}}{\textrm{T}_{13}} = 0.2 + 0.23(\textrm{v-v}_{\textrm{\textsc{LSR}}})^{2}.
\label{eq:parabola}
\end{displaymath}

As a part of the fitting process, we only use the channels with $^{12}$CO and $^{13}$CO emission above 3$\sigma$ and the minimum ratio value was fixed at zero velocity. Figure \ref{Fig:correction} shows the fit with a $\chi^{2}$ of 0.4 as a solid green line and the blue dots correspond to the weighted mean values and the error bars are the weighted standard deviations in each channel. For this fit, we did not use the last three points, presented as the red dots, because at these velocities $^{12}$CO starts becoming optically thin.

After applying the correction factors to every channel and using the emission that traces the outflow in the blue- and red shifted components, we start from the assumption that the emission is optically thin and in Local Thermodynamic Equilibrium (LTE). Next, we integrated the intensity from pixels with detections above 5$\sigma$ over all of these velocity channels to measure the N$_{\textrm{CO}}$ column density. Then, with an X$_{\textrm{co}} = 10^{-4}$, which is the abundance of CO relative to H$_{2}$ taken from \citet{Frerking1982}, we inferred the column density of N$_{\textrm{H}_{2}}$. The interested reader can find a detailed description available in Appendix C of \citet{Dunham2014}. 
This quantity is multiplied by pixel area and then, summing over all the pixels, we obtain estimates of mass (M$_{\textrm{ch}}$) and using velocity channels, estimates of momentum (M$_{\textrm{ch}}$ \textrm{v}$_{\textrm{ch}}$)\footnote{Properties not corrected from inclination effects} and energy (0.5 M$_{\textrm{ch}}$ \textrm{v}$_{\textrm{ch}}^{2}$) of the outflow. To obtain total values of these parameters, it is integrated over the whole range of velocities with detections. 

In order to avoid contamination by ambient cloud emission or material not related to the outflows of HBC 494, we integrate only over those channels with emissions that shape the outflow; for example, in Figure \ref{Fig:CanalCO21}, the $^{12}$CO emission in the channels with a velocity range between 1.75 and 7.25 km s$^{-1}$ were not considered, because they are related to the emission that arises from around the protostar and parcels of matter that belong to the surrounding envelope. Additionally, in order to assure emission only from the outflow, we built a mask around HBC 494 of radius size $\sim$ 1.5$^{"}$, where pixels inside this area were removed from the final analysis. Thus, separating the red- and blue shifted components, the blue shifted outflow kinematics were estimated by integrating channels in the range between -5.25 and 1.5  km~s$^{-1}$ for $^{12}$CO and, 1.0 and 4.0  km~s$^{-1}$ for $^{13}$CO. The range of channels in the red shifted emission are between 7.25 and 16.5  km~s$^{-1}$ for $^{12}$CO and, 4.5 and 6.25  km~s$^{-1}$ for $^{13}$CO. 

For simplicity and considering how the excitation temperature varies the estimated parameters \citep[e.g.][]{Curtis2010, Dunham2014}, we adopted for this quantity, values of 20 and 50 K in our analysis. The estimated parameters are shown in Table \ref{table:kinematics}. Additionally, taking the extent of the $^{12}$CO emission (20") and the maximum speed of the $^{12}$CO gas extension, obtained using $\frac{\rm v_{s}- v_{b}}{2}$ where $\rm v_{s}$ and $\rm v_{b}$ are the red- and blue-shifted maximum velocities, we estimated a kinematic age for HBC 494 of $\sim$5400 years to obtain the mechanical luminosity and mass loss rate of the outflow, see Table \ref{table:kinematics}. However, these property estimations are lower limits of the outflow because HBC 494 is not detected completely in the extension of 20" of the image, see e.g. Figures \ref{Fig:CO12} and \ref{Fig:CO18}, and implicit assumptions in the method.

\subsubsection{Envelope Mass and  Kinematics}

Following the process described above, we use the C$^{18}$O emission to estimate the lower limits of mass and dynamical properties of the envelope surrounding HBC 494. Taking into account that as an optically thin tracer only provides information of distant regions from the central object, we did not build a mask for the C$^{18}$O cube. Also, it is not necessary to apply a correction factor to compute the parameters of the cloud. As previously performed, we also separated the blue- and red-shifted components to integrate over ranges between and  1.75 and 4.25 km~s$^{-1}$ for blue-shift velocities and 4.5 and 6.25  km~s$^{-1}$ for red shift velocities. Estimated parameters of the molecular cloud are shown in Table \ref{table:kinematics}.

\begin{table*}
 \centering
 \begin{minipage}{128mm}
  \caption{Mass, Momentum, Luminosity and Kinetic Energy of the Outflow and Envelope}
 \label{table:kinematics}
  \begin{tabular}{clccccc}
 \hline \hline 
 \multicolumn{1}{c}{\textbf{}} &
  \multicolumn{1}{c}{\textbf{}} &
\multicolumn{2}{c}{\textbf{Blue shifted}\footnotemark[1]}    &

\multicolumn{2}{c}{\textbf{Red shifted}\footnotemark[2]}  \\[0.5ex] 

 \multicolumn{1}{c}{\textbf{Isotope}} &
 \multicolumn{1}{c}{\textbf{Property}} &
\multicolumn{1}{c}{\textbf{20 (K)} }  &
\multicolumn{1}{c}{\textbf{50 (K)}} &
\multicolumn{1}{c}{\textbf{20 (K)}}    &
\multicolumn{1}{c}{\textbf{50 (K)}}\\[0.5ex]\hline \hline \\[-3ex]  

\multirow{5}{*}{\begin{sideways}{\Large{ $^{12}$CO}} \end{sideways}}
&Mass  (10$^{-2}$ M$_{\odot}$)  &                                3.40 (52.00) & 5.03 (77.00)    &  5.54 (50.00)&   8.20 (74.00) \\ 
&Mass loss (10$^{-5}$ M$_{\odot}$ yr$^{-1}$)   &           0.64 (9.70)& 1.00 (14.32 )  & 1.03 (9.32)& 1.53 (13.80)  \\ 
&Momentum  (M$_{\odot}$ km s$^{-1}$)   &  0.20 (2.60)&  0.30 (3.90)&0.28 (1.90) &  0.40 (3.02) \\ 
&Energy (10$^{43}$ ergs.)   &                                        1.10 (13.24)& 1.60  (19.60 ) & 1.71 (9.44) & 2.53 (14.00)         \\ 
&Luminosity (10$^{-2}$L$_{\odot}$)   &                      2.00 (20.00) & 3.00 (30.00) & 3.00 (15.00)& 4.00 (22.00) \\[0.1ex]\hline

\multirow{5}{*}{\begin{sideways}{\Large{ $^{13}$CO}} \end{sideways}}
&Mass  (10$^{-2}$ M$_{\odot}$)  &  3.46 &5.20 & 3.10 & 4.70\\ 
&Mass loss ( 10$^{-6}$ M$_{\odot}$ yr$^{-1}$)  &  6.50   & 9.80    &5.81    & 8.80 \\ 
&Momentum  (10$^{-2}$ M$_{\odot}$ km s$^{-1}$) &    6.10  &9.20 &1.70 &2.60\\ 
&Energy (10$^{42}$ ergs.)   &    1.14  & 1.73   & 0.14 & 0.21 \\ 
&Luminosity (10$^{-3}$ L$_{\odot}$)   & 1.76 & 2.70   & 0.21 &  0.32  \\[0.1ex]\hline

\multirow{5}{*}{\begin{sideways}{\Large{ C$^{18}$O}} \end{sideways}}
&Mass  (10$^{-4}$ M$_{\odot}$)  &   3.78 &  5.74 & 11.91 & 18.09   \\ 
&Mass loss (10$^{-7}$M$_{\odot}$ yr$^{-1}$)   &  0.71  &  1.10 & 2.22  & 3.39   \\ 
&Momentum  (10$^{-4}$ M$_{\odot}$ km s$^{-1}$)   &  2.00    &   3.00 & 2.35 & 3.60\\ 
&Energy (10$^{39}$ ergs.)   &   1.90   &    2.90 & 0.68 & 1.04  \\ 
&Luminosity (10$^{-6}$ L$_{\odot}$)   &    2.92 & 4.44& 1.05 & 1.60\\[0.1ex]\hline
\end{tabular}
$^{1}$ Blue shifted outflow kinematics were estimated after a cut above 5$\sigma$ and integration of channels  between -5.25 and 1.5  km~s$^{-1}$ for $^{12}$CO, 1.0 and 4.0  km~s$^{-1}$ for $^{13}$CO, and  1.75 and 4.25 km~s$^{-1}$ for C$^{18}$O\\
$^{2}$ Red shifted outflow kinematics were estimated with a threshold value above 5$\sigma$ and integration of channels between 7.25 and 16.5  km~s$^{-1}$ for $^{12}$CO,  4.5 and 6.25  km~s$^{-1}$ for $^{13}$CO, and 4.5 and 6.25  km~s$^{-1}$ for C$^{18}$O.\\
$^{3}$ Parameters inside the parentheses correspond to the computed values after applying the correction factors for optical depth effects to all the channels with $^{13}$CO detection above 5$\sigma$. 
\end{minipage}
\end{table*}

\begin{figure}
    \centering
        \centering
        \includegraphics[width=0.46\textwidth]{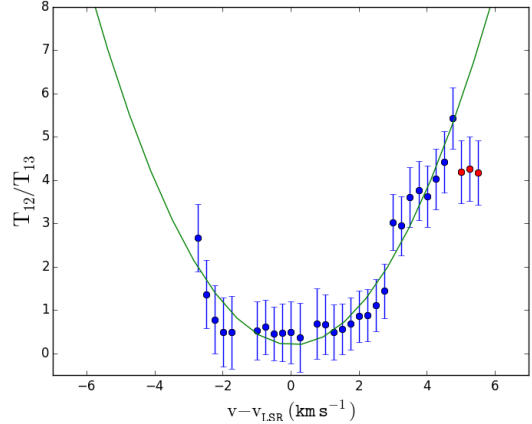}
        \caption{ Ratio of the brightness temperatures $\frac{T_{12}}{T_{13}}$ as a function of the velocity from the systemic velocity. The blue solid  dots are the weighted mean values and the error bars are the weighted standard deviations in each channel. The red solid dots are weighted mean values not used in the fitting process. The green solid line is the best-fit second-order polynomial with a $\chi^{2}$ of 0.4.}
        \label{Fig:correction}
    \end{figure}

\subsection{Spectral Lines}
\label{sec:lines}

To further explore the kinematic properties of the envelope interacting with the outflow cavities, we obtained spatially integrated $^{13}$CO and C$^{18}$O spectral lines from the regions where most of the radiation arises from the envelope while trying to avoid background emission. The regions enclosed with dashed lines in Figures \ref{Fig:CO13b} and \ref{Fig:CO18b} correspond to the regions in which we integrated the line profiles and figure \ref{Fig:spectra} shows the line profiles of both isotopes. As expected, the observed C$^{18}$O spectral line shows a typical width previously seen in quiescent envelopes of around 0.7-2 km s$^{-1}$ \citep[e.g.][]{Jorgensen2002, Kristensen2012}.

The observed C$^{18}$O line has a FWHM of 0.84  $\pm$ 0.04 and is centred at 4.60 $\pm$ 0.02 km~s$^{-1}$, which is taken as the systemic velocity. The $^{13}$CO profile shows a dip around 4.0 km~s$^{-1}$ and is slightly blue-shifted from the systemic velocity. This particular profile is not a real absorption and might be due to missing short-spacing information in the u-v coverage. The C$^{18}$O line is not affected by this spatial filtering because it is less abundant and traces a more compact region. In contrast to that, the $^{13}$CO traces a more extended and energetic envelope region, which is evident in its full width of around 4.5 km~s $^{-1}$.  Hence, the $^{13}$CO profile is broader than the C$^{18}$O indicating the complex outflow motion in $^{13}$CO, previously seen in very young objects embedded in molecular clouds \citep[e.g][]{Kristensen2012}. This profile might be evidence of an expanding envelope mostly in the Southern region, likely due to the stream emission observed with the $^{13}$CO and C$^{18}$O isotopes at the southeast of the disc. Additionally, it might also indicate how the outflow cavities are efficiently pushing aside envelope material, see Figures \ref{Fig:CO12a} and  \ref{Fig:CO13a}. For a Class I object with a low mass envelope, it becomes relatively easy to expel material to outer regions of the molecular cloud and thus, modify the geometry of the molecular cloud and surroundings \citep[e.g][]{Kristensen2012, Mottram2013}. We provide a more detailed discussion about this scenario in Section \ref{Sec:cloud}.

\begin{figure}
\centering
\includegraphics[width=0.51\textwidth]{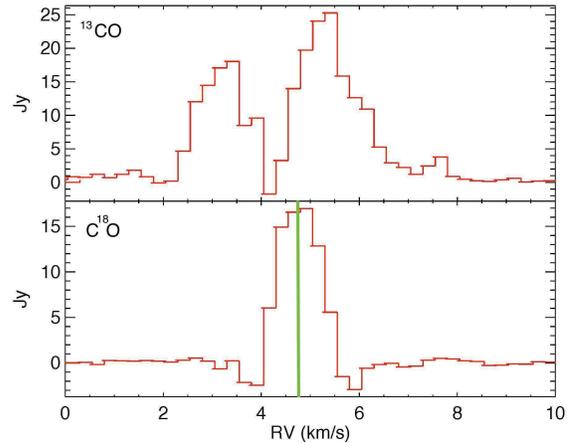}
\caption{$^{13}$CO  and C$^{18}$O  line profiles of the envelope shown as red solid lines. The spectrum for C$^{18}$O is centered at 4.60 $\pm$ 0.02 km~s$^{-1}$ and is shown as a green vertical line.}
\label{Fig:spectra}
\end{figure}

\section[]{Discussion} 
\label{Sec:Discussion}


\subsection{The Wide Angle Outflow} 
\label{Sec:wide}

Figures  \ref{Fig:CO12} and \ref{Fig:CO13grey} illustrate a double-lobed $^{12}$CO structure of HBC 494. The main characteristics of these shell-like structures are the \textit{wide}-angle outflows extending ($\sim$ 8000 au) in opposite directions with a projected opening angle of $\sim$150$^{\circ}$ and a velocity pattern with high and low velocity components along and at the wake of the lobes, see Figure \ref{Fig:CanalCO21}. This defined wide-angle morphology of the cavities has previously been observed in only a few Class I objects such as RNO 129 \citep[$\sim$125$^{^{\circ}}$;][]{Arce2006} and more recently in the FU Ori V883 Ori \citep[$\sim$150$^{^{\circ}}$;][]{RuizRodriguez2016} that, with an outflow mass on the order of 10$^{-2}$ M$_{\odot}$, are observational evidence of the evolutionary trend found in the morphology of the molecular outflows by \citet{Arce2006}. This trend basically indicates how the opening angles increase with source age, leading to a stronger outflow-envelope interaction during the evolutionary process. Additionally, in Figure \ref{Fig:CO13}, a ``stream" is observed to the southeast of HBC 494 with velocities between 1$-$1.75 km~s$^{-1}$ and partially linked to this stream, a C$^{18}$O emission is detected with a velocity range of 1.75$-$2.0 km~s$^{-1}$, see Figure \ref{Fig:CO18a}. The direction of this stream likely follows the blue-shifted outflow component on this side of the bipolar cone and might be related to flows that arise from the interaction of the highly accreting disc inner edges with an existing threading strong magnetic field \citep[e.g.][]{Donati2005}. It has been suggested that outflows can be centrifugally accelerated along net vertical magnetic field lines threading the disc; when non-ideal magnetohydrodynamical effects are taken into account, the MRI turbulent activity disappears, allowing that the disc launches a strong magnetocentrifugal wind \citep{Bai2013}. If so, the wind sweeps up the ambient molecular gas in the vicinity of the surrounding envelope when it interacts with a collapsing environment \citep{Snell1980, Shu2000, Gardiner2003, Bjerkeli2016}. Figure \ref{Fig:art} depicts a computer illustration of HBC 494 to better show the dynamic properties of the system components.  The effectiveness of the removal of envelope material depends on the age of the protostar and the degree of collimation of the wind. Thus, a highly collimated wind with high density profile would flow along the rotation axis (e.g. Class 0), while for a more evolved protostar with a reduction of envelope material, the collimated wind would decrease with its density at increasing angles from the axis \citep{Shang1998}. For a Class I object, the gas entrained by the wide-angle component of the wind will be the dominant structure in the molecular outflow, producing the observed widening angle outflows seen in HBC 494. 

Also, it is expected that the energetic outflows impact an homogeneous surrounding envelope, because the bipolar extension are largely symmetric in their opening-angles. Nevertheless, as mentioned previously the envelope material interacting with the outflow in the Northern region of HBC 494 might be slightly denser than in the Southern region, see Figures \ref{Fig:CO12a} and \ref{Fig:CO12b}. Figure \ref{Fig:projection} shows an image of the Re50N nebula taken as part of The Two Micron All Sky Survey \citep{Skrutskie2006} in H band (1.6 $\micron$), where the continuum and $^{12}$CO contours are over-plotted indicating the position of HBC 494 and its wide outflows.  From this image, the inhomogeneous molecular cloud hosts the outflow evolution with an embedded outflow emerging at the Southern side, while the Northern cavity is still deeply embedded. However, it is difficult to confirm this from our estimates of the outflow parameters, see Table \ref{table:kinematics}, because are subject to the maximum recoverable scale and the spatial filtering, e.g. Figure \ref{Fig:spectra}. Therefore, these values are underestimated and biased by the visibility sampling, which require observations for a larger scale structure. Although, the parent molecular cloud might not be homogeneously distributed around HBC 494,  this does not mean that both lobes would evolve in a distinguishable way and might vary only in the extension of their outflows. Unfortunately, information with respect to the extension projected onto the rotation axis is more ambiguous due to the observing limitations, which did not detect the complete structure of the outflow cavities. In addition, we might be facing a very young binary object as the triggering mechanism of the outburst \citep{Cieza2016}, and considering that FUor outbursts are usually accompanied by strong winds \citep[e.g][]{Bastian1985} and if most of the stellar mass is accreted during these events, we speculate that the origin of the wide-angle cavities of HBC 494 is a combination of binarity and magnetocentrifugally-driven winds.

\subsection{Role of Outflows in Star Formation }
 Inspecting Table \ref{table:kinematics}, the parameters estimated from the corrected  $^{12}$CO emissions are increased by a factor ranging from $\sim$ 10 to 20. This supports the claim by \citet{Dunham2014} that molecular outflows are much more massive and energetic than commonly reported.  Our mass estimations from the blue- and red-shifted emissions, after correcting for optical depth, are on the order of outflows previously presented using a similar approach \citep[e.g][]{Arce2006, Curtis2010, Dunham2014}. If we are indeed facing the fact that outflows are more massive than expected, and from a theoretical point of view, these outflows are responsible for extracting angular momentum from the proto-stellar core-disc system to allow the accretion onto the central object, then we might expect a huge impact on the final stellar mass and indirectly, the shape of the Initial Mass Function (IMF). It has been suggested by theoretical models that including outflow feedback the average stellar mass would decrease by a factor of $\sim$ 3, while the number of stars increases by a factor of $\sim$ 1.5 \citep{Federrath2014}. Energetic outflows may be responsible for triggering star formation by perturbing other regions of the same cloud, which leads to the formation of multiple stars instead of a single one \citep{Federrath2014}. 
 
Additionally, we note that the outflow mass, momentum, and energy for HBC 494 are higher than those found recently for the Class I objects, HL Tau and V2775 Ori \citep{Klaassen2016, Zurlo2016}. However, these quantities highly depend on the outflow mass and in the case of HL Tau and V2775 Ori, these values were not corrected for optical depth effects.  On the other hand, HL Tau and V2775 Ori present a kinematic age of 2600 yr with opening angles of 90$^{^{\circ}}$ and 30$^{^{\circ}}$, respectively, while HBC 494 has a kinematic age of $\sim$5400 yr and an opening angle of 150$^{^{\circ}}$. For both HL Tau and V2775 Ori the opening angle differs considerably and these objects, differ in their kinematic age from HBC 494. In spite of the fact that the outflows present a significant discrepancy in their spatial extension, it highlights the importance of understanding and constraining the initial conditions of the stellar formation process, which might be controlled by the parent molecular cloud and perturbations from external agents to develop this variety of FU Ori objects.

\subsection{Role of Outflows in Disc Evolution}

The physical source generating the wide angle outflow in HBC 494 could be connected to the evolution of the disc and as a requirement, it must play an important role in the removal of angular momentum from the accreted material. Indeed, it has been indicated that outflowing gas from an accretion disc might provide an efficient transport of angular momentum to permit the accretion of matter onto the central star \citep{Blandford1982}. In addition, to determine the disc lifetime it is necessary to constrain properties such as mass accretion and mass loss rates, which might be intrinsically coupled to angular momentum transport \citep{Gressel2015, Bai2016}. In recent years, the ratio of the wind mass loss rate to the wind driven accretion has been suggested to be on the order of 0.1 with large uncertainties \citep[e.g][]{Klaassen2013, Watson2015}. As shown in \citet{Dunham2014}, and from HBC 494 as well, the mass loss rates are underestimated, leading to suggest that the mass loss rate is not a small fraction of the accretion rate. Then,  if increasing the mass outflow by factors of $\sim$ 20 or even higher \citep[50-90 in the most extreme cases;][]{Dunham2014} , it might limit the accretion directly by factors of similar magnitude. Recently, \citet{Bai2016a} suggested a scenario where discs might lose mass at a considerable fraction of the accretion rate. Enhancing the FUV penetration depth would lead to considerable increases of the outflow rates, while the accretion rate would also present an increase in a more moderate way \citep{Bai2016}. Using the relation $\dot{M}=1.25 \frac{L_{acc}R_{\star }}{GM_{\star} }$ \citep{Hartmann1998} and if we consider HBC 494 as a Class I star of 1 $\rm \Msun$ with a 250 $\rm \Lsun$ \citep{Reipurth1986}, 3 $\Rsun$ \citep{Baraffe2015} and assuming an age of 0.5 Myrs, typical of Class I objects, then its accretion rate is $\sim$ 3$\rm x$10$^{-5}$ $\rm \Msun$ $ \rm y^{-1}$, which is of the order of typical FU Ori objects. From Table \ref{table:kinematics}, the estimates for mass loss rates are of the order of the accretion rate, however, these values might be taken with caution because our approach to estimate the accretion rate might failed in the limits of very large accretion luminosities \citep{Hartmann1998}. More importantly, the rate of mass outflow and angular momentum transport increase with increasing net vertical magnetic field, however, we are still limited to accurately predict the timescale of disc evolution due to the lack of knowledge of magnetic flux distribution threading the disc \citep{Bai2013, Bai2016}.

As mentioned above, our outflow mass and kinematic estimates of HBC 494 are massive and energetic, meaning that in order to raise these massive outflows via winds might require a fully ionised inner disc. For a normal classical T Tauri star, it has been shown that the low ionisation levels in the disc prevent the rise of winds from the inner region. The wind outflow rate sensitively depends on the FUV and X-ray penetration depth which on the other hand are determined by the abundance of small grains \citep{Bai2016} and the photon energy \citep{Ercolano2013}. For FUors, in the region interior to the dust sublimation radius, the disk is sufficiently ionised so that the magnetocentrifugal wind can be launched efficiently. A very important consequence of this strong penetration depth and hence, the wind mass loss, is that mass is primarily removed from the disc surface as a function of the distance \citep{Bai2016}. This creates favorable conditions for planetesimal formation in outer regions of the disc, where most dust grains settle at the disc midplane. Recently, from high resolution ALMA observations, \citet{Cieza2016a} reported the detection of the water snow line at a distance of $\sim$ 42 au from the central star, V883 Ori. They suggested that the location of the water snow-line at these early stages is largely affected by outburst events. In addition, observational evidence has been found for preferential loss of gas relative to dust in CO isotopologue surveys by \citet{Williams2014} and \citet{Ansdell2016}. Thus, the disc evolution and initial conditions for planetary formation at these early stages might be strongly influenced by the intimately connected processes of outflows and accretion \citep{Bai2016a}.

 \subsection{Role of Outflows in Star-Forming Molecular Clouds}
 \label{Sec:cloud}

Class I objects are able to disperse their surrounding envelope effectively, through their short outburst events. Thus, the outflows could energetically expel a large amount of material from the molecular cloud modifying neighbouring structures \citep{Federrath2014}. A possible change of structure in the neighbouring regions of HBC 494 could have origin in a previous outburst event. As previously mentioned, HBC 494 seems to start digging out from the molecular cloud, exposing its Southern side and thus, expelling material more efficiently to large distances. It was noticed by \citet{Chiang2015} that Re50 located at the south of Re50N (1.5 arc min, associated nebula to HBC 494), started fading to the west and moved eastwards by a curtain of obscuring material, while the Re50N suffered a dramatic increase in brightness sometime between 2006 and 2014 \citep{Chiang2015}. Considering the direction of the observed $^{12}$CO emission in HBC494 and making a projection of the emerging outflows, it coincides with the direction of the expulsion of material directed towards the northeast side of Re50 \citep[see Figure 2 in][]{Chiang2015}. Therefore, material that was possibly pushed by a HBC 494 outflow could be responsible for the significant decrease in brightness of this neighbouring object.

\begin{figure}
\includegraphics[width=0.48\textwidth]{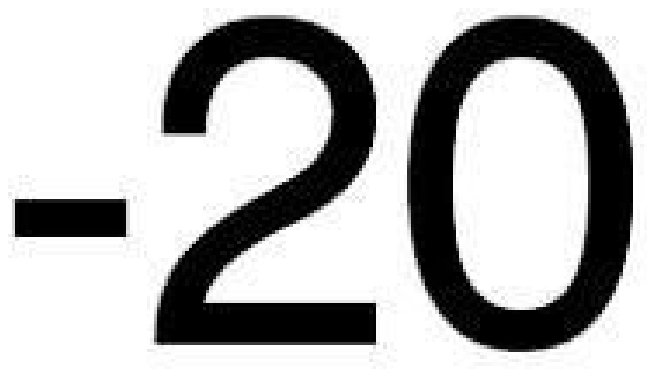}
\caption{Projection of the blue and red $^{12}$CO emission delineating the outflows over the H-band \citep[1.6 $\micron$;][]{Skrutskie2006} image of HBC 494. Blue and magenta contours show the integrated intensity of the $^{12}$CO blue- and red shifted lobes, respectively, at 80, 150, 300, 450 and 600 $\times$ 3$\sigma$ levels. The green contours are the continuum emission and represent the position of HBC 494.}
\label{Fig:projection}
\end{figure}

\section[]{Summary}
\label{Sec:Summary}

We studied the HBC 494 system using $^{12}$CO, $^{13}$CO and C$^{18}$O images at 0.2$''$ resolution.  With a large expansion of the molecular outflows ($\sim$ 8000 au), the likely non-uniformity of the envelope material of HBC 494 lead us to suggest that the evolution of the outflows is largely influenced by a differential density and degree of interaction between outflow and surrounding envelope on both sides of the bipolar cone. This scenario might be a result of the binarity and magnetocentrifugally-driven wind present in the system  \citep[e.g][]{Bastian1985, Terquem1999}. Although, higher-resolution images are needed to confirm whether HBC 494 is a close binary like FU Ori itself \citep{Hales2015}.  Using the $^{12}$CO, $^{13}$CO and C$^{18}$O emissions, we derived the mass and kinematics of the outflow on the order of 10$^{-2}$ M$_{\odot}$ for the mass and 10$^{-3}$ M$_{\odot}$ km s$^{-1}$ for the momentum. After correcting for optical effects, these properties increased by a factor ranging from 10-20. This increase in the kinematic properties might be observational evidence of the important role played by the outflows in FU Ori objets to drive the evolution of the disc via winds and hence, the conservation of angular momentum \citep{Bai2016}.

\section*{Acknowledgments}

We thank Yichen Zhang for helpful discussions to improve the quality of this work.  L.A.C., S.C., D.P.,  S.P. and A.Z.  acknowledge support from the Millennium Science Initiative (Chilean Ministry of Economy),  through grant Nucleus  RC130007. L.A.C. was also supported by CONICYT-FONDECYT grant number 1140109. D.P. was also supported by FONDECYT grant number 3150550. H.C. acknowledges support from the Spanish Ministerio de Econom\'ia y Competitividad under grant AYA2014-55840P. KM acknowledges the support by the Joint Committee of ESO and Government of Chile.

This paper makes use of the following ALMA data:
ADS/JAO.ALMA No. 2013.1.00710.S . ALMA is a partnership of
ESO (representing its member states), NSF (USA) and NINS (Japan),
together with NRC (Canada), NSC and ASIAA (Taiwan), and
KASI (Republic of Korea), in cooperation with the Republic of
Chile. The Joint ALMA Observatory is operated by ESO, AUI/NRAO
and NAOJ. The National Radio Astronomy Observatory is a facility
of the National Science Foundation operated under cooperative
agreement by Associated Universities, Inc.

This publication makes use of data products from the Two Micron All Sky Survey, which is a joint project of the University of Massachusetts and the Infrared Processing and Analysis Center/California Institute of Technology, funded by the National Aeronautics and Space Administration and the National Science Foundation.

\bibliographystyle{mn2e} 
\bibliography{biblio}

\bsp	
\label{lastpage}

\end{document}